\documentclass[acmsmall]{acmart}

\AtBeginDocument{%
  }

\usepackage{booktabs}
\usepackage{multirow}
\usepackage{tabularx}
\usepackage{amsmath}

\usepackage{amssymb}
\usepackage{xspace}
\usepackage{enumitem}
\usepackage{subcaption}
\usepackage{pifont}
\usepackage{xcolor}
\usepackage{bbold}
\usepackage{framed}

\newcommand{\system}{\textsc{SALUS}\xspace}
\newcommand{\minibench}{\textsc{BIRD-Clean-xs}\xspace}
\newcommand{\bird}{\textsc{BIRD}\xspace}
\newcommand{\spider}{\textsc{Spider}\xspace}

\newcommand{\cmark}{\ding{51}}
\newcommand{\xmark}{\ding{55}}

\newcommand{\finding}[2]{%
  \vspace{4pt}%
  \noindent\hspace{0.03\columnwidth}%
  \colorlet{shadecolor}{gray!15}%
  \begin{minipage}{0.93\columnwidth}%
    \begin{shaded}%
      \textbf{Finding #1.}\enspace{\itshape #2}%
    \end{shaded}%
  \end{minipage}%
  \vspace{4pt}%
}

\hypersetup{hidelinks}

\title[SALUS: Automated Auditing of NL-to-SQL Benchmarks]{SALUS: Automated Auditing of NL-to-SQL Benchmarks through Weak Supervision of Multi-Agent Output}

\author{Shiyuan Zhou}
\email{szhou20@uci.edu}
\affiliation{%
  \institution{University of California, Irvine}
  \city{Irvine}
  \state{California}
  \country{USA}
}

\author{Ashwin Gerard Colaco}
\email{acolaco@uci.edu}
\affiliation{%
  \institution{University of California, Irvine}
  \city{Irvine}
  \state{California}
  \country{USA}
}

\author{Sainyam Galhotra}
\email{sg@cs.cornell.edu}
\affiliation{%
  \institution{Cornell University}
  \city{Ithaca}
  \state{New York}
  \country{USA}
}

\author{Sharad Mehrotra}
\email{sharad@ics.uci.edu}
\affiliation{%
  \institution{University of California, Irvine}
  \city{Irvine}
  \state{California}
  \country{USA}
}

\hypersetup{pdfauthor={Shiyuan Zhou, Ashwin Gerard Colaco, Sainyam Galhotra, Sharad Mehrotra}}

\begin{document}

\setcopyright{cc}
\setcctype{by}
\acmJournal{PACMMOD}
\acmYear{2026}
\acmVolume{4}
\acmNumber{6 (SIGMOD)}
\acmArticle{374}
\acmMonth{12}
\acmPrice{}
\acmDOI{10.1145/3856369}

\received{17 April 2026}
\received[revised]{19 August 2026}
\received[accepted]{12 September 2026}



\begin{abstract}

Natural language to SQL (NL-to-SQL) benchmarks are foundational to progress in data analysis research,
yet recent work has shown that widely-used benchmarks contain significant annotation errors. These errors silently corrupt evaluation metrics, penalize correct model
output, and distort the field's understanding of state-of-the-art performance. We present \system\footnote{The source code is available at
\url{https://github.com/ucisharadlab/SALUS}.}, a system that automatically detects annotation errors in NL-to-SQL benchmarks. \system frames benchmark auditing as a weakly supervised error detection: SQL generated by multiple LLM agents drive a suite of complementary weak-labeling functions. By passing this noisy vote matrix through a generative label model, we extract high-confidence training samples without requiring human ground truth. These samples train a decision plane that maps gold SQL query features to per-agent trustworthiness, allowing \system to intelligently fuse reliability estimates with raw verdicts for rigorous benchmark error detection. We evaluate on \minibench, a benchmark of 298 \bird development tasks with manually verified correctness labels. \system achieves F1\,=\,0.9194, significantly outperforming the state-of-the-art baselines. Applying \system to the full development sets, we estimate annotation error rates of approximately \ignorespaces {37\%} on \bird and \ignorespaces {27\%} on Spider.
\end{abstract}

\maketitle

\section{Introduction}\label{sec:introduction}

NL-to-SQL systems have become foundational for data analytics, business intelligence, and database interfaces. Benchmarks such as \bird~\cite{li2023bird}, \spider~\cite{yu2018spider}, and WikiSQL~\cite{wikisql} play a central role in this ecosystem: they define standardized evaluation protocols, determine leaderboard rankings, and shape research priorities. Reported accuracy improvements drive claims of state-of-the-art performance and influence both academic research and industrial adoption.

\begin{figure}[t]
    \centering
    \includegraphics[width=0.80\columnwidth]{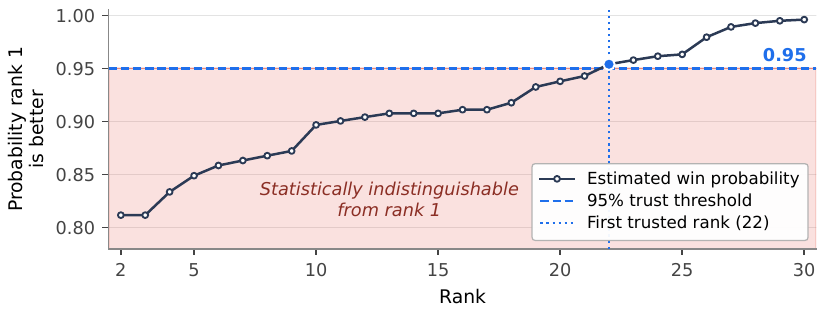}
    \caption{Probability that the rank-1 system on the \bird development set truly outperforms lower-ranked methods under the estimated \ignorespaces {37\%} annotation error rate. The rank-1 system cannot be reliably distinguished from the \ignorespaces {next 20}.}
    \Description{The plot compares the estimated probability that the rank-1 BIRD development-set system outperforms each lower-ranked system with a 95 percent threshold. Under the estimated 37 percent annotation error rate, comparisons with the next 20 systems do not reach that threshold.}
    \label{fig:bird_leaderboard_trust}
\end{figure}

These leaderboards, however, are built on a fragile assumption: that the benchmark annotations (commonly called ``gold'' annotations) are correct. The \bird leaderboard separates its top systems by only 1--3 percentage points of execution accuracy, yet when we manually audited 200 random queries we found that 34\% of the gold annotations are incorrect. When the noise in a benchmark exceeds the gap between competing systems, the leaderboard no longer measures what it claims. As shown in Figure~\ref{fig:bird_leaderboard_trust}: under our estimated \ignorespaces {37\%} error rate, the rank-1 system on the \bird development set cannot be reliably distinguished from the \ignorespaces {next 20}.

This is not an isolated finding. Independent audits of \bird and \spider have reported error rates ranging from 7\% to over 50\%~\cite{jin2025pervasive, yang2025sqldriller, liu2025nl2sqlbugs}. Nor is it solved by manual review: the \bird team conducted an expert re-annotation campaign in November 2025~\cite{li2023bird}, yet 47 of the 68 errors in our random sample persisted unchanged in the updated release. The errors are numerous and diverse, spanning incorrect SQL logic, ambiguous natural-language questions, and inconsistent data files, and they resist manual correction because human annotators face the same challenges of domain complexity and fatigue that introduced the errors in the first place.

Automating detection is itself challenging, not only for NL-to-SQL benchmarks but for any setting where reference annotations may be unreliable. No oracle labels exist: the gold annotations are themselves the objects under assessment, so standard supervised learning does not apply. The errors are heterogeneous, so no single heuristic can reliably capture their diversity. Furthermore, any LLM-based detection signal is itself fallible: agent-gold disagreement does not guarantee an annotation error, nor does agreement preclude one, and agents may fail entirely by producing unexecutable SQL.

Existing methods address some of these challenges but not all. Hybrid approaches such as SAPAR~\cite{jin2025pervasive} generate LLM diagnostic reports but rely on human expert review for the final decision, inheriting the limits of manual annotation. SQLDriller~\cite{yang2025sqldriller}, the most comparable fully automated method, assumes the natural-language question and schema are correct and generates 10 structurally diverse SQL candidates per query. It flags an error only when at least one candidate outperforms the gold SQL, restricting its coverage to pure SQL errors and scaling LLM calls to compensate for candidate failure. Our key observation is different: no single LLM agent is an oracle, but each carries a diagnostic signal that can be extracted and combined. When multiple frontier models independently produce mutually consistent SQL that contradicts the gold annotation, it becomes increasingly unlikely that all models are wrong in the same way.

We introduce \system, an automated and benchmark-agnostic framework that converts noisy multi-agent disagreement into reliable benchmark audits through weak supervision. Given each benchmark task, \system prompts $K$ independent LLM agents to produce candidate SQL queries. It extracts execution, structural, and intent-alignment signals from these outputs, encoding them as diverse labeling functions (LFs) distributed across complementary signal sources so that each captures failure modes invisible to the others. A generative label model aggregates these noisy votes into probabilistic correctness scores, yielding high-confidence training samples without human labels. These samples train per-agent reliability models conditioned on gold SQL structural features, learning which agents to trust for which query types rather than weighting reliable agents and fallible agents equally. A final ensemble combines these reliability estimates with raw agent verdicts to produce the error-detection label.

We evaluated \system on \minibench, a ground-truth benchmark of 298 tasks for the detection of annotation errors in \bird with manually verified labels. Among all fully automatic baselines, \system achieves the best performance with an F1 of 0.9194, outperforming the 4-agent majority vote (F1\,=\,0.8895){, the automated verifier stage of SAPAR (F1\,=\,0.8389),} and SQLDriller (F1\,=\,0.6456). When applied at full scale, \system estimates annotation error rates of \ignorespaces {37\%} in the \bird development set and \ignorespaces {27\%} in the \spider development set. Our perturbation analysis shows that for any pair of methods whose development-set accuracy gap falls within \ignorespaces {5.09\%}, their performance difference cannot be reliably distinguished at the 95\% confidence level. Rank correlation analysis further indicates that the hidden test set may share similar annotation artifacts, limiting the ability of the current leaderboard to reliably distinguish top-ranked systems.

We summarize our contributions as follows:

\begin{itemize}[leftmargin=*, topsep=0pt]
    \item We propose \system, a benchmark-agnostic weak supervision framework that automatically detects annotation errors using multi-agent output. While we demonstrate it on NL-to-SQL benchmarks, the underlying approach applies to any domain where reference annotations may be unreliable.

    \item We construct \minibench, a manually labeled ground-truth benchmark of 298 tasks for \bird annotation error detection with a five-category error taxonomy, combining errors corrected by the \bird team with random samples. \system achieves F1\,=\,0.9194 on \minibench, the highest among all comparable methods.

    \item We apply \system to the full \bird and \spider development sets, estimating annotation error rates of approximately \ignorespaces {37\%} in \bird and \ignorespaces {27\%} in \spider, and show that this benchmark noise propagates directly into leaderboard rankings, rendering score gaps between top systems statistically indistinguishable.
\end{itemize}

\section{Related Work}\label{sec:related_work}

\subsection{NL-to-SQL Benchmarks and Evaluation}\label{sec:rw_benchmarks}

NL-to-SQL systems translate natural-language questions into executable SQL queries, enabling
users to interact with relational databases. To measure the effectiveness of these systems,
several benchmarks have been developed, evolving from simple single-table lookups to complex
reasoning tasks across diverse domains. WikiSQL~\cite{wikisql} was an early large-scale
effort focused on single-table queries, while Spider~\cite{yu2018spider} introduced the requirement
for cross-domain generalization with multi-table schemas and complex joins. More recently, Spider~2.0~\cite{spider2} and BIRD~\cite{li2023bird} have emerged as primary benchmarks for realistic NL-to-SQL tasks, emphasizing large real-world databases and complex SQL generation under practical constraints and external evidence.

A central design choice in NL-to-SQL benchmarks is how to evaluate SQL answers. Testing SQL for benchmarking is different from using SQL for data analysis. In real analysis, multiple SQL formulations may all be acceptable if they provide useful insight. However, in a benchmark, each NL question must be paired with a fixed reference answer so that predictions can be deterministically judged and compared fairly across systems. To support this goal, prior work has used two main metrics: Exact match accuracy~\cite{yu2018spider}, strictly compares structural syntax but rigidly penalizes valid, semantically equivalent queries. Consequently, modern benchmarks prioritize the second metric, execution accuracy, which compares executed result sets on the database instance to better capture true semantic equivalence. However, this also means that the gold annotation must itself be correct: otherwise, correct system outputs may be penalized, and erroneous ones may be rewarded. Benchmark quality is therefore essential for meaningful evaluation.

Current leaderboards report execution accuracies above 80\% on BIRD and 90\% on Spider~2.0 but these numbers are computed against gold annotations whose correctness is assumed but not guaranteed. Our manual investigation of 200 random samples from the BIRD development set revealed an error rate of 34\%, suggesting that the metrics reported may substantially misrepresent the true performance of the model.

\subsection{Benchmark Quality and Error Detection}\label{sec:rw_quality}

Manual annotation auditing is the most naive way to detect benchmark errors. It requires domain expertise, is prohibitively time-consuming, does not scale to modern benchmarks with thousands of instances, and still cannot avoid human annotation errors. Several recent studies have therefore developed automated or hybrid methods to detect errors in NL-to-SQL benchmarks.
NL2SQL-BUGs~\cite{liu2025nl2sqlbugs} proposed a two-level error taxonomy (9 categories, 31
subcategories) and prompted LLMs with the taxonomy followed by manual examination to evaluate
the benchmark. Jin et al.~\cite{jin2025pervasive} built SAPAR, an LLM-based pipeline where
the iteratively called LLM agent generates verification queries and diagnostic reports to be reviewed by a human expert. SQLDriller~\cite{yang2025sqldriller} introduced execution consistency as a
detection criterion: an LLM agent generates structurally diverse SQL candidates, uses a formal SQL equivalence checker to craft synthetic database instances that distinguish non-equivalent queries, and prompts an LLM to execute the natural language question on these instances. Each SQL is scored by counting execution-result matches with the LLM's natural-language execution results. Mappings for which gold SQL scores lower than an alternative are flagged as likely incorrect.

These efforts have confirmed that annotation errors are prevalent in frontier benchmarks.
NL2SQL-BUGs detected 106 errors in the BIRD development set (6.91\% of their sample);
\citet{yang2025sqldriller} manually found 810 (37.7\%) errors in the Spider test set and 768
(50.1\%) in the BIRD development set. \citet{jin2025pervasive} reported error rates of 52.8\% in
BIRD Mini-Dev and 62.8\% in Spider~2.0-Snow, with corrections shifting leaderboard rankings by
up to nine positions. The wide variance in reported rates reflects differences in methodology and error definitions, but the consistent finding is that benchmark noise is systematic and consequential.
The BIRD team has acknowledged the problem: in November 2025, they released an updated development set~\cite{li2023bird} produced by a quality review program led by five PhD students and industry engineers, incorporating community feedback collected over the preceding year. While this effort corrected a number of annotations, our manual investigation found that 47 of the 68 errors in our 200 random samples persisted unchanged in the updated release.

These limitations motivate the need for a fully automated benchmark auditing technique that can
detect annotation errors without human intervention and be reapplied whenever datasets are updated. \system addresses this gap: rather than relying on extensive LLM calls or requiring manual expert review, it aggregates multiple noisy agentic detection signals through a weak supervision workflow and assigns each query a final detection label, enabling accurate, scalable, and repeatable auditing of NL-to-SQL benchmarks. 

We emphasize that annotation error detection and annotation correction are related but distinct tasks. Reliable detection is a prerequisite for correction: proposing a trustworthy replacement SQL first requires identifying that the original annotation is likely wrong, whereas benchmark auditing only requires reliable detection of problematic items. For this reason, \system focuses on detection, which is sufficient to extract a clean evaluation subset from noisy benchmarks. Prior work, such as SQLDriller couples error detection with the generation of candidate corrections. However, our results on \minibench indicate that, for this underlying detection task, SQLDriller performs substantially worse than \system (F1 = 0.6456 versus 0.9194), underscoring the need for more advanced auditing techniques. Nonetheless, the design of a robust, fully automatic correction mechanism remains an important direction for future research, which we further elaborate on in Section~\ref{sec:discussion_correction}.

\subsection{Weak Supervision}

Weak supervision replaces hand-labeled ground truth by aggregating multiple imperfect programmatic labeling sources. Domain knowledge is encoded as \emph{labeling functions} (LFs), such as heuristics, pattern matchers, or pretrained models, that emit noisy labels. A generative label model estimates the LFs' accuracies from their agreement patterns without ground truth, producing probabilistic labels to train a downstream discriminative model~\cite{ratner2017snorkel}. This paradigm scales robustly in practice~\cite{bach2019snorkeldrybellcasestudy} and can match fully supervised approaches requiring thousands of hand-labeled examples~\cite{boxwrench2024neurips,zhang2021wrench}.

Annotation error detection in NL-to-SQL benchmarks is a natural fit for this paradigm
for two reasons. First, no ground-truth correctness labels exist: determining whether a
gold SQL annotation is correct requires expensive expert review, and the scale of modern
benchmarks makes exhaustive auditing impractical. Second, the available evidence is
heterogeneous and complementary: execution result consistency, SQL structural analysis,
and natural-language intent alignment each capture a different facet of annotation
quality, yet none is individually reliable enough to determine correctness. Accordingly, \system designs LFs to encode that evidence and provide noisy votes which a generative label model aggregates into probabilistic correctness scores for the downstream pipeline. Our experiments on \minibench, a 298-task benchmark with manually verified correctness labels (Section~\ref{sec:minibench}), confirm this design: cross-category LF pairs exhibit a mean Pearson correlation of only 0.081, and removing any single signal-source category degrades overall detection F1 by up to 0.0436 (Section~\ref{sec:system_eval}).

\section{\system}\label{sec:system}

\begin{table}[t]
\centering
\caption{Summary of notation.}
\label{tab:notation}
\small
\setlength{\tabcolsep}{4pt}
\begin{tabular}{@{}cl@{}}
\toprule
\textbf{Symbol} & \textbf{Description} \\
\midrule
$\mathcal{B}, N, i$ & Benchmark of $N$ tasks, indexed by $i$ \\
$(q_i, s_i, D_i)$ & NL question, gold SQL, and database for task $i$ \\
$R(s, D)$ & Result set from executing $s$ on database $D$ \\
$\mathcal{A}, K, j, A_j$ & Set of $K$ LLM agents, indexed by $j$ \\
$\hat{s}_{i,j}, v_{i,j}$ & Candidate SQL and verdict of $A_j$ on task $i$ \\
$m_{i,j}$ & Validity mask of $A_j$ on task $i$ \\
$M$ & Number of labeling functions \\
$\Lambda$ & Vote matrix $\in \{\textsc{Correct}, \textsc{Incorrect}, \textsc{Abstain}\}^{N \times M}$ \\

$\hat{y}_i$ & Hard weak label for task $i$ \\
$\mathbf{x}_i, V$ & $V$-dimensional SQL feature vector $\mathbf{x}$ for task $i$ \\
$t_{i,j}, \hat{r}_j(\mathbf{x}_i)$ & Trust target and predicted reliability of $A_j$ \\
$fb_i, p_i^{fb}$ & Fallback hard label and probability for task $i$ \\
\bottomrule
\end{tabular}
\end{table}

\subsection{Task Formulation}\label{sec:task_formulation}

Table~\ref{tab:notation} summarizes the notation used throughout this section. An NL-to-SQL benchmark consists of $N$ tasks $\mathcal{B} = \{(q_i, s_i, D_i)\}_{i=1}^{N}$, where $q_i$ is a natural-language question, $s_i$ is the gold SQL annotation, and $D_i$ is the associated database. An annotation error exists when $s_i$ does not correctly capture the intent of $q_i$. The goal is to predict a binary correctness label for each task without access to oracle labels.

To obtain a detection signal, \system prompts $K$ independent LLM agents $\mathcal{A} = \{A_1, \dots, A_K\}$, each producing a candidate query $\hat{s}_{i,j}$ given the database schema and NL question. A binary verdict $v_{i,j}$ records whether agent $A_j$'s execution result matches the gold annotation on $D_i$. We define a per-agent validity mask $m_{i,j} \in \{0,1\}$, where $m_{i,j} = 1$ indicates that $A_j$ produced a query on task $i$ that can execute on $D_i$ without any compile-time error. Only valid agents contribute to downstream training and aggregation.

\begin{figure*}[htbp]
    \centering
    \includegraphics[width=\linewidth]{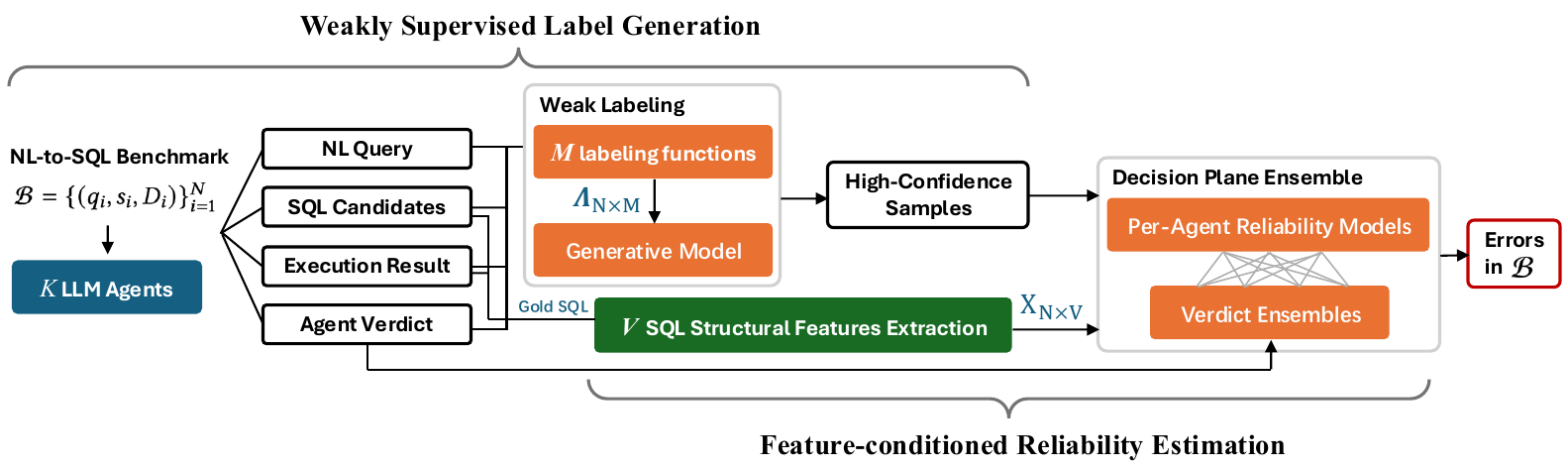}
    \caption{An overview of the \system pipeline.}
    \Description{Benchmark tasks are sent to multiple LLM agents. The question, candidate SQL, execution results, and agent verdicts supply labeling functions whose vote matrix is aggregated by a generative model. High-confidence samples train per-agent reliability models. SQL structural features and agent verdicts enter the decision plane ensemble, which predicts benchmark annotation errors.}
    \label{fig:workflow}
\end{figure*}

When all $K$ agents produce SQL queries whose execution results agree with each other but disagree with the gold annotation, their consensus provides evidence of a possible annotation error. Under an illustrative independence assumption, if each agent produces the correct answer with probability $p$, the probability that all $K$ agents are wrong and nevertheless agree is at most $(1-p)^K$; with $K=4$ and $p=0.65$, this bound is approximately 1.5\%. Multi-agent disagreement signals are informative, but agents are not equally reliable across all queries: one may excel on queries with subqueries but struggle with complex joins, while another exhibits the opposite pattern. \system therefore learns \emph{where} each agent is trustworthy and weights their contributions accordingly, through three components (Figure~\ref{fig:workflow}): weakly supervised label generation (Section~\ref{sec:weak_labeling}), feature-conditioned reliability estimation (Section~\ref{sec:reliability}), and ensemble aggregation (Section~\ref{sec:ensemble}). The specific system configurations used in our evaluation are detailed in Section~\ref{sec:experimental_setup}.

\subsection{Weakly Supervised Label Generation}\label{sec:weak_labeling}

To maximize independence of failure modes, agents are drawn from distinct model families and providers, reducing the risk of correlated errors~\cite{wisdom_crowd}. Each agent receives an identical prompt containing the database schema, the natural-language question, any provided evidence, and is instructed to return only the SQL query.

From these agent outputs, we extract a set of $M$ labeling functions (LFs) across three orthogonal signal sources, where each LF emits a vote in $\{\textsc{Correct},\; \textsc{Incorrect},\; \textsc{Abstain}\}$ for every query. The first are the \emph{Execution signals} that compare result sets between agent and gold queries, detecting direct output disagreements. \emph{Structural signals} compare SQL at the Abstract Syntax Tree (AST) level, catching errors such as missing joins or aggregation mismatches even when execution results coincidentally agree.
Finally, \emph{Intent signals} check whether the SQL reflects salient cues in the natural-language question, such as whether ``how many'' maps to \texttt{COUNT} or ``top $N$'' maps to \texttt{LIMIT}.

Within each source, LFs vary in polarity: bipolar functions emit both \textsc{Correct} and \textsc{Incorrect} votes depending on the observed condition, while unipolar functions vote in only one direction and abstain otherwise. Applying all $M$ LFs to every task produces the vote matrix $\Lambda \in \{\textsc{Correct},\; \textsc{Incorrect},\; \textsc{Abstain}\}^{N \times M}$. The specific LF instantiation is detailed in Section~\ref{sec:lf_config} and Appendix~\ref{app:lf}.

Because no oracle labels exist, we aggregate the $M$ noisy votes through a generative label model~\cite{ratner2017snorkel} rather than relying on any single function. The model learns the accuracy and correlation structure of each LF from the observed vote matrix $\Lambda$ alone, without assuming that all LFs are equally reliable. LFs that frequently agree with the majority receive higher estimated accuracy, while noisy or low-coverage LFs are down-weighted. The model then combines these learned accuracies to produce a per-task probabilistic correctness score. Using that score, we construct a high-confidence training set by selecting the same number of top-ranked tasks from each predicted class and assign each a hard weak label $\hat{y}_i \in \{0, 1\}$. These weak labels are the only supervision used in the downstream components; no human annotations are consulted during training or model selection.

\subsection{Feature-Conditioned Reliability Estimation}\label{sec:reliability}

\begin{figure}[t]
    \centering
    \includegraphics[width=0.70\columnwidth]{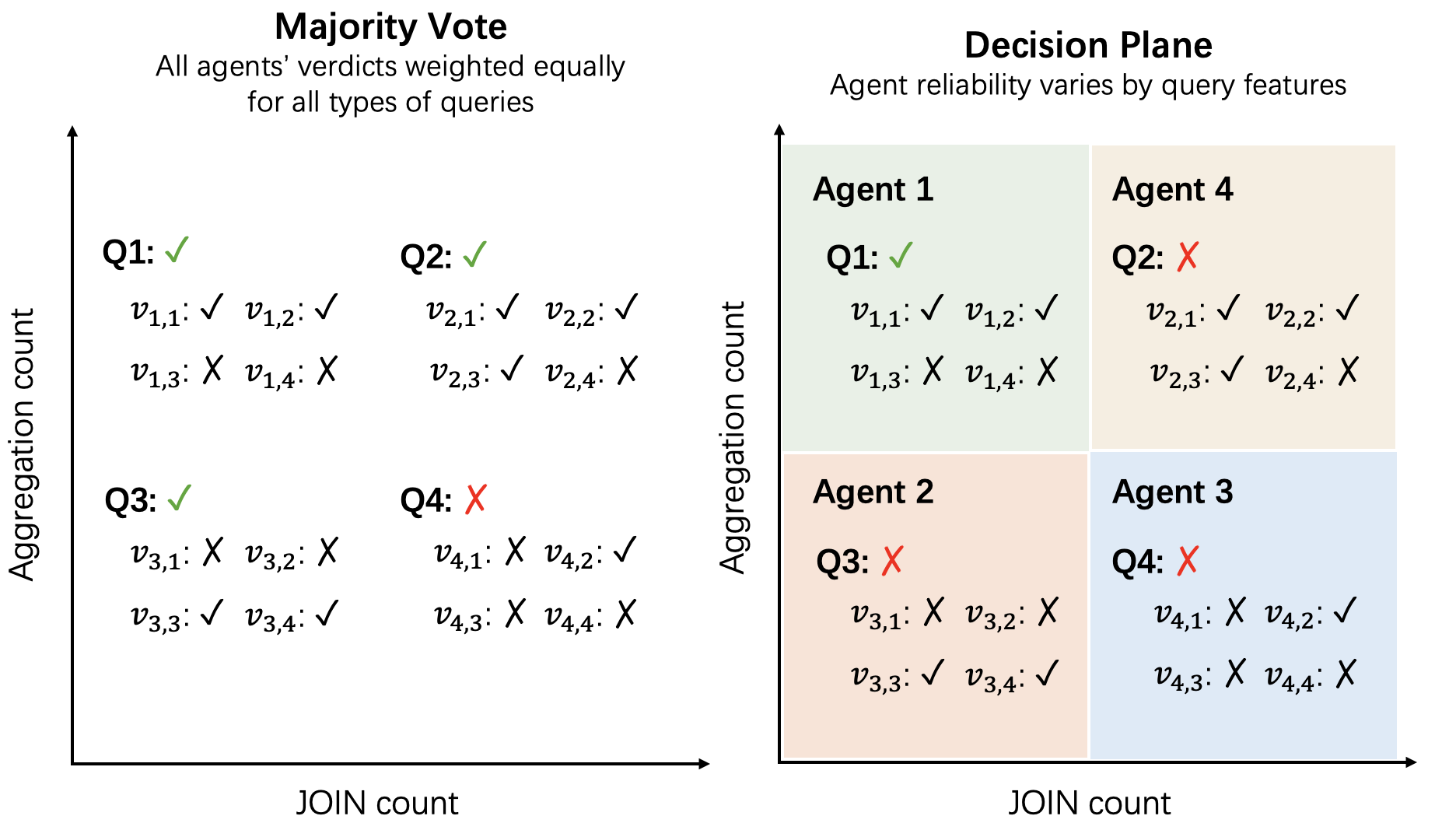}
    \caption{Conceptual demonstration of the comparison of majority vote and the decision plane. 
    Each query (Q1--Q4) shows four agent verdicts where \cmark{} = agent execution matches gold, \xmark{} = agent execution 
    differs from gold). \textbf{Left:} majority vote weights all agents 
    equally and predicts \textsc{Incorrect} only when more than half of 
    valid agents disagree with the gold SQL. \textbf{Right:} the decision 
    plane assigns each query to a region of the SQL feature space where a 
    specific agent is most reliable (indicated by the region label and 
    background color), and follows that agent's verdict.}
    \Description{Two panels compare predictions for four queries over join count and aggregation count. Majority voting predicts Q1, Q2, and Q3 correct and Q4 incorrect. Feature-conditioned routing divides this space into four regions assigned to different agents and predicts Q1 correct and Q2, Q3, and Q4 incorrect.}
    \label{fig:decision_plane}
\end{figure}

The high-confidence samples from Section~\ref{sec:weak_labeling} provide training signal, but applying a single global rule to all queries discards useful structure. As conceptually illustrated in Figure~\ref{fig:decision_plane}, an agent reliable on queries with low join count may struggle when aggregation count increases, and vice versa. We capture this through what we call the \emph{decision plane}: each gold SQL $s_i$ and its execution result map to a feature vector $\mathbf{x}_i \in \mathbb{R}^{V}$ encoding structural properties. The system learns a separate reliability function $\hat{r}_j(\mathbf{x}_i)$ for each agent $A_j$ over this space, so that queries with similar $\mathbf{x}_i$ are routed to the agent most trustworthy in that part of the feature space.

These features are agent-agnostic and fall into three groups: structural flags indicating the presence of constructs such as \texttt{HAVING}, \texttt{DISTINCT}, \texttt{GROUP BY}, and nested subqueries; count features capturing the number of joins, tables, aggregates, and \texttt{WHERE} conditions; and execution features including result cardinality and whether the gold SQL raises an error. The specific dimensionality is detailed in Section~\ref{sec:feature_config}.

For each agent $A_j$, we define a binary trust target $t_{i,j} = \mathbb{1}[v_{i,j} = \hat{y}_i]$, indicating whether the agent's verdict is consistent with the weak label on task $i$. We then train a separate reliability model on high-confidence samples, restricted to tasks where $m_{i,j}=1$: $\hat{r}_j(\mathbf{x}_i) = P(t_{i,j}=1 \mid \mathbf{x}_i)$,
where $\hat{r}_j(\mathbf{x}_i)$ is a continuous trust score representing the predicted probability that agent $A_j$ is reliable on a query with features $\mathbf{x}_i$. A separate task-level fallback model is trained directly on $(\mathbf{x}_i \to \hat{y}_i)$ over high-confidence samples, producing a hard correctness label $fb_i$ and a probability $p_i^{fb} = P(\hat{y}_i = 1 \mid \mathbf{x}_i)$. This model provides predictions for queries where no agent produced valid SQL and supplies a continuous task-level signal to the ensemble.

\subsection{Ensemble Aggregation}\label{sec:ensemble}

The final stage combines reliability estimates with raw agent verdicts to predict each task's correctness label. 
\emph{Weighted voting} computes a validity-masked, reliability-weighted average, using notation in Table~\ref{tab:notation}:
\[
\hat{wv}_i =
\begin{cases}
\mathbb{1}\!\left[
\dfrac{\sum_j m_{i,j}\,\hat{r}_j(\mathbf{x}_i)\, v_{i,j}}
{\sum_j m_{i,j}\,\hat{r}_j(\mathbf{x}_i)}
\geq 0.5
\right], & \text{if } \sum_j m_{i,j}>0, \\[6pt]
fb_i, & \text{otherwise.}
\end{cases}
\]

\begin{figure*}[t]
    \centering
    \includegraphics[width=\linewidth]{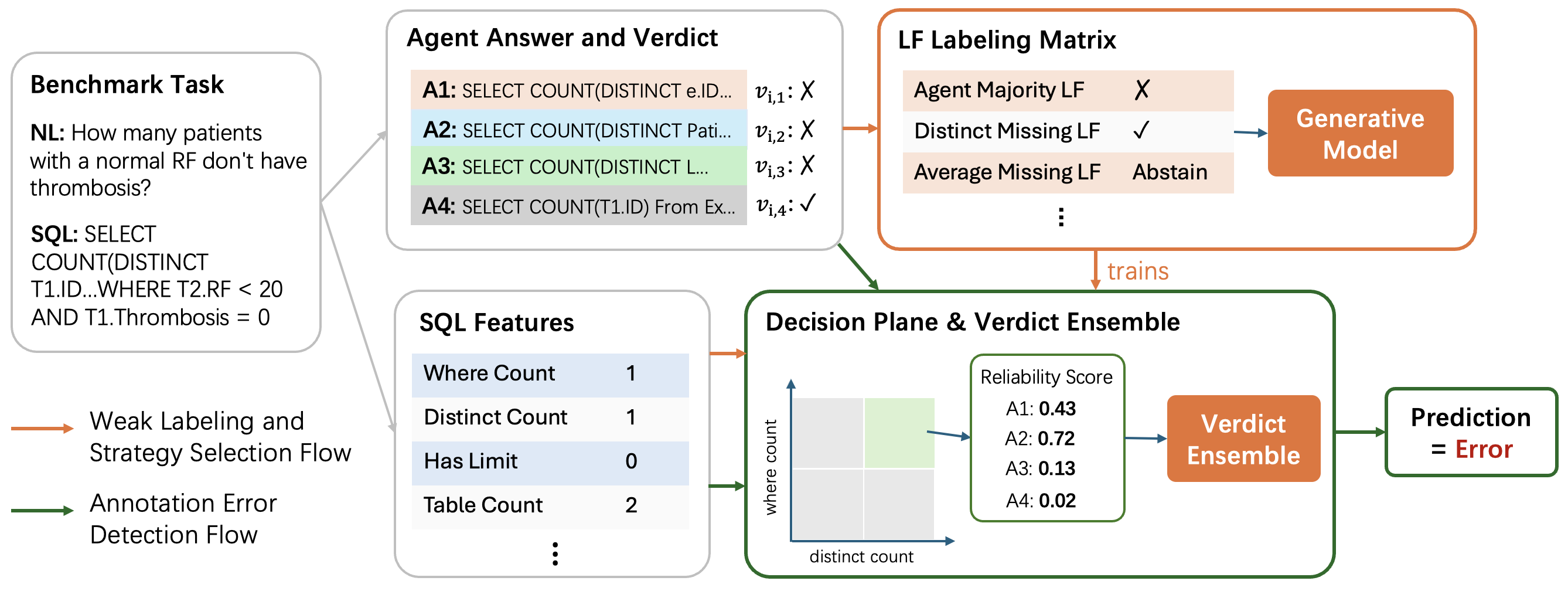}
    \caption{End-to-end example of \system on a single benchmark task. 
    The orange path shows the training flow: agent verdicts and gold SQL 
    produce labeling function votes, which a generative model aggregates 
    into high-confidence labels that supervise the decision plane. The 
    green path shows inference: SQL structural features map the query to 
    a decision plane region, producing per-agent reliability scores that 
    are combined with agent verdicts to predict the final label. The 
    visualized decision plane is a conceptual 2D projection; the actual 
    model operates over $V$ features.}
    \Description{A patient-count query illustrates the pipeline. Three agents disagree with the reference and one agrees. Example labeling functions vote incorrect, correct, or abstain. The label model provides weak supervision. SQL features include where count 1, distinct count 1, no limit, and table count 2. The decision plane assigns agent reliability scores of 0.43, 0.72, 0.13, and 0.02; the ensemble predicts an annotation error.}
    \label{fig:intermediate_values}
\end{figure*}

\emph{Top-$k$ voting}, which selects the $k$ most reliable agents and applies majority vote within that subset; a \emph{stacked meta-learner} that trains an XGBoost~\cite{chen2016xgboost} classifier on the concatenation of reliability scores, verdicts, and validity masks; and \emph{score fusion}, which trains a logistic regression on the weighted voting score and the fallback probability $p_i^{fb}$, incorporating the task-level signal as a continuous feature for every query rather than only as a zero-valid safety net. The reliability-model family and aggregation strategy are jointly selected by 5-fold stratified cross-validation on high-confidence samples, optimizing F1 for the error class. After selection, the strategy is retrained on full high-confidence set and applied to all benchmark tasks to determine the final detection label using their corresponding agent verdicts and the value of the SQL features. Figure~\ref{fig:intermediate_values} illustrates the entire pipeline on a single benchmark task, showing how \system generates intermediate values for model training and final prediction.

\section{\minibench}\label{sec:minibench}

\begin{table*}[t]
\centering
\caption{Representative examples of each annotation error type in \minibench, with the number of errors per category.}
\label{tab:error_examples}
\small
\setlength{\tabcolsep}{4pt}
\begin{tabularx}{\linewidth}{@{}>{\raggedright\arraybackslash}p{0.18\linewidth}r>{\footnotesize\raggedright\arraybackslash}X>{\footnotesize\raggedright\arraybackslash}X@{}}
\toprule
\textbf{Error Type} & \textbf{\# Errors} & {\small\textbf{Question}} & {\small\textbf{Explanation}} \\
\midrule
Incorrect SQL &
134 &
List out the account numbers of female clients who are oldest and has lowest average salary\ldots &
Gold SQL appends \texttt{LIMIT 1}, returning only one account when multiple accounts satisfy the condition. \\
\addlinespace
Query Ambiguity &
14 &
Which active district has the highest average score in Reading? &
``Average score'' can refer to a single school's score or the mean across all schools in a district; the gold SQL selects a single school rather than aggregating by district. \\
\addlinespace
Data Inconsistency &
7 &
How many posts were created on 21st July, 2010? &
Two tables record creation dates for the same posts but contain different row counts; querying different tables yields different results. \\
\addlinespace
Misinterpreting Domain Semantics &
8 &
Provide the hair colour of the human superhero who is 185\,cm tall. &
The question assumes a unique answer, but multiple human superheroes are 185\,cm tall, so the gold SQL returns more than one result. \\
\addlinespace
Incorrect Evidence &
3 &
State the race and year of race in which Michael Schumacher had his fastest lap. &
The provided evidence references ``Alex Yoong'' instead of Michael Schumacher, introducing a factual error into the annotation context. \\
\midrule
\textbf{Total} & \textbf{166} & & \\
\bottomrule
\end{tabularx}
\end{table*}

\subsection{Overview}\label{sec:GT_overview}

\minibench is our primary benchmark for evaluating NL-to-SQL annotation error detection. It contains 298 tasks with manually verified labels: 166 erroneous annotations and
132 correct ones. All tasks are drawn from the \bird development set used in SQLDriller repository \cite{yang2025sqldriller} for better comparison between two methods. This version is slightly different with the original published development set by BIRD team by 103 query changes with 44 in gold SQL, 30 in natural-language question, and 38 in evidence.

\minibench is constructed from the union of two complementary sources. The first is a 200-task random sample from
the full development set. The second is a targeted set of 109 confirmed errors identified from
version differences between two \bird development set. All 298 tasks are manually verified.

\subsection{Source 1: Random Sampled Set}\label{sec:GT_RS}

We randomly sampled 200 tasks from the development set ($n=1{,}534$) and manually reviewed each one, yielding 68 erroneous tasks and 132 correct ones (34.0\% error rate). To confirm
that the sample is representative of the full development set, we performed chi-square tests on the
distributions of difficulty level and database category between the sample and the population, with
Bonferroni correction at family-wise $\alpha=0.05$. Neither difficulty ($p=0.124$) nor belonged database
($p=0.999$) show a significant deviation, supporting that the sample is unbiased.

Of the 68 confirmed errors, 21 were subsequently changed by the \bird team in their November 2025
update, while the other 47 were left unchanged. Thus, at least 47 confirmed errors persisted through the update, showing that a substantial fraction remained unaddressed even after the \bird team's own manual review. Among the tasks the BIRD team did change, the majority involved SQL query modifications, though some changes modified the natural-language question or evidence.

Because this subset carries manually verified ground-truth labels on a representative sample, the known error rate
together with the system's evaluated precision and recall on this subset enables a confidence
interval estimate for the true benchmark-wide error rate
(Section~\ref{sec:true_error}).

\subsection{Source 2: Version-Diff Candidate Set}\label{sec:GT_diff}

This set exploits a natural signal from benchmark maintenance. In November 2025, the \bird team
released an updated development set, indicating that they were actively fixing known annotation
issues. We do not assume the updated release is uniformly correct; rather, we use the differences
between the earlier BIRD development set used by SQLDriller and the November 2025 release to identify tasks that are especially likely to be incorrect
in the earlier version.

Starting from the two versions, we identify tasks whose gold SQL changed, yielding 475 candidate
tasks. We then apply two filters: (1) we retain only tasks whose natural-language question and
evidence remain unchanged, so that the difference is attributable to the SQL annotation itself; and
(2) we execute both SQL versions on the underlying database and retain only tasks whose execution
results differ. After filtering, 109 candidate tasks remain. Manual examination confirms all 109 as
erroneous in the previous version.

\subsection{Labeling Process}\label{sec:labeling}

All tasks in \minibench are manually verified. Two PhD students independently reviewed each
NL-to-SQL pair, judging whether the gold SQL correctly answers the natural-language question.
{Before adjudication, the two annotators agreed on 86.24\% of the 298 tasks (Cohen's $\kappa \approx 0.73$, substantial agreement).}
Disagreements were escalated to two supervising professors with expertise in database systems, who
made the final determination.

To further improve labeling quality, for all tasks in \minibench, we cross-checked our judgments against external evidence,
including NL2SQL-Bugs~\cite{liu2025nl2sqlbugs}, Arcwise's reported \bird
errors~\cite{arcwise}, and issue reports from the official \bird
repository~\cite{BIRD_panel}. Whenever our judgment disagreed with an external source, we
conducted an additional re-validation round before finalizing the label.

\subsection{Error Taxonomy and Statistics}\label{sec:error_stats}

During manual labeling, we classified each confirmed error into one of five categories.
\textit{Incorrect SQL} denotes cases where the gold SQL does not correctly answer the
natural-language question. \textit{Query Ambiguity} refers to cases where the natural-language
question admits multiple valid interpretations and the gold SQL captures only one.
\textit{Data Inconsistency} refers to cases where columns intended to share the same meaning across
tables contain inconsistent values, so different but defensible queries yield different results.
\textit{Misinterpreting Domain Semantics} refers to cases where the query relies on implicit domain
assumptions that do not hold in the actual data.
\textit{Incorrect Evidence} refers to cases where the provided evidence or hints contain factual
errors that lead to a wrong SQL annotation.
Table~\ref{tab:error_examples} summarizes the count for each error type across \minibench, and provides a representative example of each category.

\textit{Incorrect SQL} is the dominant category, accounting for 134 of 166 errors (80.7\%). The remaining 32 errors are distributed across \textit{Query Ambiguity} (14), \textit{Data Inconsistency} (7), \textit{Misinterpreting Domain Semantics} (8), and \textit{Incorrect Evidence} (3). While direct SQL errors dominate, a non-trivial minority arise from other types of errors (19.3\%), underscoring the need for auditing tools capable of covering diverse error types.

\section{Evaluation}\label{sec:evaluation}

We evaluate \system along four axes: end-to-end detection accuracy on \minibench (Section~\ref{sec:detection_minibench}), head-to-head comparison with \ignorespaces {existing auditing systems, including their cost and latency} (Section~\ref{sec:sqldriller}), calibrated error rate estimation across NL-to-SQL benchmarks (Section~\ref{sec:true_error}), and the impact of annotation noise on leaderboard trustworthiness (Section~\ref{sec:leaderboard}). We then analyze the pipeline components to demonstrate the system design (Section~\ref{sec:system_eval}).

\subsection{Experimental Setup}\label{sec:experimental_setup}

\paragraph{Agent Configurations.}
We evaluate \system under two agent configurations where $K=4$. \textbf{Frontier}: GPT-5, GPT-4o, Claude Opus~4.6, and Gemini~2.5~Pro. This is our primary configuration and represents the frontier LLMs. \textbf{\ignorespaces {Legacy}}: GPT-4o-mini, GPT-4o, Claude~3~Haiku, and Haiku~4.5. This configuration more closely matches the model generation available when \ignorespaces {the prior auditing systems SQLDriller and SAPAR were published}, enabling a fairer comparison to prior work. Each agent receives the schema, question, and evidence for one SQL-generation attempt. {For both configurations, reasoning and thinking parameters remain unset, preserving provider-default behavior.}

\paragraph{Weak Labeling Configuration.}\label{sec:lf_config}
We instantiate the weak labeling framework (Section~\ref{sec:weak_labeling}) with 43 task-level LFs and 1 meta LF ($M = 44$ total). Table~\ref{tab:lf_summary} reports the breakdown by signal source, category, and polarity. Detailed LF design is in Appendix~\ref{app:lf}. The generative label model is instantiated with the Snorkel label model~\cite{ratner2017snorkel}. We select the top 250 high-confidence samples for each label, yielding a balanced training set of 500 tasks.

\begin{table}[t]
\caption{Breakdown of labeling functions by signal source, category, and label polarity.}
\label{tab:lf_summary}
\centering
\small
\begin{tabular}{@{}llrc@{}}
\toprule
Signal Source & Category & Count & Polarity (B/U) \\
\midrule
\multirow{2}{*}{Execution}
  & Result-set comparison     & 14 & 5\,/\,9 \\
  & Agent consensus           &  2 & 2\,/\,0 \\
\midrule
\multirow{3}{*}{Structural}
  & AST comparison            &  9 & 3\,/\,6 \\
  & Schema consistency        &  1 & 0\,/\,1 \\
  & Known error patterns      &  6 & 0\,/\,6 \\
\midrule
Intent
  & NL--SQL alignment         & 11 & 3\,/\,8 \\
\midrule
Aggregate
  & Meta (vote-vector)        &  1 & 1\,/\,0 \\
\midrule
  & \textbf{Total}            & \textbf{44} & \textbf{14\,/\,30} \\
\bottomrule
\end{tabular}
\end{table}

\paragraph{SQL Feature Configuration.}\label{sec:feature_config}
We extract $V = 12$ features as described in Section~\ref{sec:reliability}. Structural flags comprise 6 binary features: \texttt{DISTINCT}, \texttt{GROUP BY}, \texttt{HAVING}, \texttt{ORDER BY}, \texttt{LIMIT}, and nested subquery. Count features comprise 4 numeric features: join count, table count, aggregate function count, and \texttt{WHERE} clause count. Execution features comprise 2 features: result cardinality and a binary execution error indicator.

\paragraph{Reliability Model Configuration.}\label{sec:dp_config}
Per-agent reliability models and the fallback task model (Section~\ref{sec:reliability}) are selected from three candidate families: depth-2 Classification and Regression Tree (CART)~\cite{cart}, depth-4 CART~\cite{cart}, and XGBoost~\cite{chen2016xgboost}. {The family is selected by 5-fold cross-validation on the high-confidence samples alone, with ties resolved in favor of the lower-capacity model for simplicity. The fallback model uses the same family as the per-agent models to keep the cross-validation search space tractable. Section~\ref{sec:feature_analysis} reports how detection quality varies across the three families.}

{\paragraph{Zero-Overlap Configuration.}
\system's default setting is transductive: it constructs weak pseudo-labels for a fixed unlabeled benchmark, selects the highest-confidence samples for downstream training, and predicts correctness across that same benchmark. Manual correctness labels are consulted only afterward to evaluate the fixed model, never during feature construction, sample selection, or training. However, weak labels assigned to evaluation tasks can be selected among the high-confidence samples used for downstream training. We therefore add comparison-only ``Zero-Overlap'' configurations as an inductive robustness test: before high-confidence selection, we exclude every evaluation task and retrain the decision plane and stacking mapping separately for each evaluation set. This tests whether downstream performance depends on memorizing the weak labels of evaluated tasks.}

\subsubsection{Baselines.}\label{sec:baselines}
We compare \system against \ignorespaces {baseline methods}.

\paragraph{Individual agents.}
Each agent serves as a standalone error detector. A task is predicted erroneous when the agent's execution result disagrees with the gold SQL execution result. If the agent did not have a valid executable SQL answer, the agent automatically failed on that query task.

\paragraph{Majority vote.}
Labels a query as 'incorrect' only if more than half of the valid agents disagreed with reference SQL; otherwise, it returns 'correct'. If no agent produced a valid execution result, the majority vote does not return a prediction.

\paragraph{Weak labels.}
The generative label model in \system produces a probabilistic
correctness score for each task's gold annotation from the labeling
function vote matrix $\Lambda$. This baseline thresholds these scores
at 0.5 to obtain hard predictions for all $N$ tasks, bypassing
feature-conditioned reliability estimation and ensemble aggregation.

\paragraph{Agent stacking.}
\ignorespaces 
{Agent stacking is an alternative design of our system: it depends on \system's LFs and Snorkel label model and replaces only the SQL-feature-conditioned decision plane with one global rule over agent verdicts. For each task, the $K=4$ agents produce a four-bit verdict vector, with 1 denoting that the agent's valid execution agrees with the reference SQL execution and 0 otherwise. The agents produce $2^K=16$ possible verdict vectors. A global rule maps each vector to either correct or incorrect, so there are $2^{16}=65{,}536$ mappings. We score every mapping against the same 500 high-confidence weak labels used for \system's downstream training and retain the maximum-F1 mapping for each configuration. Because the selected mapping sees no SQL features, it gives every task with the same verdict vector the same prediction.}

{\paragraph{Existing Methods.}
We compare with SQLDriller and SAR-Agent. SQLDriller~\cite{yang2025sqldriller} is a fully automatic benchmark auditing method based on SQL-differences. We use the system verdict on the BIRD development set from the released artifact. Prior work SAPAR~\cite{jin2025pervasive} consists of two components: an automated reviewer agent, SAR-Agent (o3 backbone), which generates a per-task correctness verdict and diagnostic report, followed by an expert human review stage based on that report. Because \system requires no manual effort, we compare with SAR-Agent only. Detailed comparisons are provided in Section~\ref{sec:sqldriller}.}

\paragraph{Agent oracle.}
We report an oracle upper bound over the same four-agent set. For each task, the oracle predicts the ground-truth label correctly whenever at least one agent's binary verdict matches that label; only tasks on which all four agents are wrong remain unsolved. This measures the maximum recoverable signal present in the current agent pool.

\subsubsection{Metrics { and Evaluation Data}.}\label{sec:metrics}
We report precision, recall, and F1 for detecting erroneous gold annotations on BIRD-Clean-xs. For benchmark-wide auditing, we report the calibrated error rate and 95\% confidence interval, {using manually verified 200-task random samples from each audited benchmark version.} For system-level analysis, we report component-level descriptive statistics.

\subsection{Detection Performance}\label{sec:detection_minibench}

We evaluate end-to-end error-detection performance of \system on \minibench (298 tasks: 166 errors, 132 correct). Table~\ref{tab:detection_metrics}(A) compares it with its principal baselines. \system achieves F1\,=\,0.9194, exceeding agent stacking (0.9021), weak labels (0.8896), majority vote (0.8895), and the best individual agent (Opus~4.6, 0.8616). Notably, weak labels alone already outperform majority vote and all individual agents, demonstrating that the complementary LF design captures richer evidence than raw agent agreement verdicts.

\begin{table}[t]
\centering
\caption{Detection performance on (A) \minibench and (B) the 200-task random sample.}
\label{tab:detection_metrics}
\label{tab:detection_rs200}
\small
\setlength{\tabcolsep}{3pt}
\begin{tabular*}{\linewidth}{@{\extracolsep{\fill}}lcccccc@{}}
\toprule
& \multicolumn{3}{c}{\textbf{(A) \minibench}} & \multicolumn{3}{c}{\textbf{(B) 200-task random sample}} \\
\cmidrule(lr){2-4}\cmidrule(l){5-7}
\textbf{Method} & \textbf{F1} & \textbf{Precision} & \textbf{Recall} & \textbf{F1} & \textbf{Precision} & \textbf{Recall} \\
\midrule

\multicolumn{7}{@{}l}{{\textit{Direct methods}}} \\
Majority Vote                   & 0.8895 & 0.8596 & 0.9217 & 0.7843 & 0.7059 & 0.8824 \\
Best Single Agent (Opus~4.6)    & 0.8616 & 0.9013 & 0.8253 & 0.7612 & 0.7727 & 0.7500 \\
\addlinespace[1pt]
\multicolumn{7}{@{}l}{{\textit{Weak-labeling methods (transductive)}}} \\
\textbf{\system} & \textbf{0.9194} & \textbf{0.9112} & \textbf{0.9277} & \textbf{0.8552} & \textbf{0.8052} & \textbf{0.9118} \\
{\system (Agent Stacking)}              & 0.9021 & 0.8889 & 0.9157 & 0.8243 & 0.7625 & 0.8971 \\
{\system (Weak Labels)}              & 0.8896 & 0.8817 & 0.8976 & 0.8027 & 0.7468 & 0.8676 \\
\addlinespace[1pt]
\multicolumn{7}{@{}l}{{\textit{Weak-labeling methods (inductive)}}} \\
{\textbf{\system (Zero-Overlap)}} & {\textbf{0.8961}} & {\textbf{0.8830}} & {\textbf{0.9096}} & {\textbf{0.8219}} & {\textbf{0.7692}} & {\textbf{0.8824}} \\
{\system (Agent Stacking \& Zero-Overlap)} & {0.8805} & {0.8531} & {0.9096} & {0.7792} & {0.6977} & {0.8824} \\
\bottomrule
\end{tabular*}
\end{table}

\paragraph{Random-sample validation.}
To better generalize system performance to the overall benchmark, we further evaluate \system on the 200-task random sample (Section~\ref{sec:GT_RS}). Table~\ref{tab:detection_rs200}(B) reports the results. \system achieves F1\,=\,0.8552, outperforming agent stacking (0.8243), weak labels (0.8027), majority vote (0.7843), and the best single agent (Opus~4.6, 0.7612). The consistent ranking across both \ignorespaces sets confirms that the gains are not specific to \minibench.

{\paragraph{Zero-overlap comparison.}
Under ``Zero-Overlap,'' \system remains the strongest non-oracle method, reaching F1 of 0.8961 on \minibench and 0.8219 on the BIRD 200-task sample, versus 0.8805 and 0.7792 for agent stacking. \system remains above weak labels and majority vote on both sets, while stacking falls below both. This suggests that the global stacking rule relies more on verdict combinations in training examples, whereas SQL-feature-conditioned reliability transfers to unseen tasks.}

\paragraph{Statistical significance.}
We use the 200-task representative subset (Section~\ref{sec:GT_RS}) to conduct a paired bootstrap test~\cite{koehn-2004-statistical}. In $10{,}000$ iterations,
we sample 200 task indices with replacement and compute F1 for \system and each baseline on the same resampled set. Table~\ref{tab:bootstrap} reports the observed $\Delta$F1, the 95\% percentile confidence interval, and the one-sided $p$-value.

\begin{table}[t]
\centering
\caption{Paired bootstrap significance test ($b=10{,}000$)}
\label{tab:bootstrap}
\small
\begin{tabular}{lccc}
\toprule
\textbf{Baseline} & $\Delta$\textbf{F1} & \textbf{95\% CI} & $p$ \\
\midrule
GPT-4o           & +0.154 & [0.095, 0.221] & $<$0.001 \\
Gemini~2.5~Pro   & +0.153 & [0.087, 0.225] & $<$0.001 \\
GPT-5            & +0.097 & [0.042, 0.156] & $<$0.001 \\
Opus~4.6         & +0.094 & [0.025, 0.167] & 0.002 \\
Majority Vote    & +0.071 & [0.035, 0.114] & $<$0.001 \\
Weak Labels   & +0.053 & [0.009, 0.100] & 0.009 \\
Agent Stacking   & +0.031 & [0.000, 0.065] & 0.028 \\
\bottomrule
\end{tabular}
\end{table}

The decision plane achieves a statistically significant improvement over all four individual agents and majority vote ($p < 0.025$, CI strictly above zero). Against weak labels, the improvement of 0.053 F1 is also significant ($p = 0.009$), confirming that the decision plane ensemble does not simply echo its heuristic training signals. Instead, it learns generalizable relationships between agent reliability and SQL structure that improve detection. Against agent stacking, the 95\% confidence interval approaches zero. \ignorespaces  {However, because agent stacking relies more heavily on weak labels, \system remains more robust than pure stacking.} Across all comparisons, the consistently positive CIs confirm that \system's improvements are not marginal artifacts of sampling variation but reflect a more effective mechanism for NL-to-SQL benchmark error detection.

\begin{table}[t]
\centering
\caption{Per-error-type detection of \system on \minibench{, with exact (Clopper--Pearson) 95\% binomial CIs on recall}.}
\label{tab:error_type}
\footnotesize
\setlength{\tabcolsep}{2.5pt}
\begin{tabular}{lcccc}
\toprule
\textbf{Error Type} & \textbf{Errors} & \textbf{Detected} & \textbf{Recall} & {\textbf{95\% CI}} \\
\midrule
Incorrect SQL                    & 134 & 127  & 0.9478 & {[0.895, 0.979]} \\
Query Ambiguity                  &  14 &  14 & 1.0000 & {[0.768, 1.000]} \\
Misinterpreting Domain Semantics             &   8 &   6 & 0.7500 & {[0.349, 0.968]} \\
Data Inconsistency               &   7 &   4  & 0.5714 & {[0.184, 0.901]} \\
Incorrect Evidence               &   3 &   3 & 1.0000 & {[0.292, 1.000]} \\
\midrule
{All errors} & {166} & {154} & {0.9277} & {[0.877, 0.962]} \\
\bottomrule
\end{tabular}
\end{table}

\begin{table}[t]
\centering
\caption{Per-difficulty detection of \system on \minibench, using difficulty labels provided by the BIRD team.}
\label{tab:difficulty}
\small
\begin{tabular}{lccc}
\toprule
\textbf{Difficulty} & \textbf{Errors} & \textbf{Detected} & \textbf{F1} \\
\midrule
Simple     & 90 & 83 & 0.9274 \\
Moderate   & 58 & 54 & 0.9076 \\
Challenging & 18 & 17 & 0.9189 \\
\bottomrule
\end{tabular}
\end{table}

\paragraph{Robustness across error types and difficulty.}
Tables~\ref{tab:error_type} and~\ref{tab:difficulty} break down \system's detection performance by error category and \bird difficulty tier. \system achieves perfect recall on \emph{Query Ambiguity} (14/14) and \emph{Incorrect Evidence} (3/3), where ambiguous questions naturally elicit divergent agent interpretations and evidence errors are surfaced when agents ignore the faulty hint. \emph{Incorrect SQL} is detected at high recall (127/134). The hardest categories are \emph{Data Inconsistency} (4/7) and \emph{Misinterpreting Domain Semantics} (6/8), where all agents execute against the same flawed data or lack the implicit domain knowledge needed to produce a distinguishing alternative, leaving little disagreement signal for the pipeline to exploit. {The exact binomial CIs in Table~\ref{tab:error_type} are accordingly wide for those rarer error families but tight for the overall recall (92.8\%, [87.7\%, 96.2\%]). The rarity of these families is itself a property of the released benchmark: blatant mistakes are the easiest for maintainers to catch, so surviving errors concentrate in \emph{Incorrect SQL} and categories requiring deeper reasoning.} Furthermore, we use the difficulty labels provided by the BIRD team: simple, moderate, and challenging. According to Table~\ref{tab:difficulty}, \system remains stable with F1 above 0.9 in every difficulty tier, showing that the method generalizes across SQL complexity rather than concentrating its gains on a narrow subset.

\finding{1}{{\system leads in both settings (transductive and inductive) on both evaluation sets, outperforming agent stacking and all other methods. The results also demonstrate the benefit of weak supervision: weak labels consistently outperform the direct methods, namely the best single agent and majority vote.}}

\subsection{Comparison with \ignorespaces  {Existing Auditing Systems}}
\label{sec:sqldriller}

NL-to-SQL annotation error detection methods fall into two different paradigms. \textbf{Execution-based methods} such as \system compare SQL execution results on the benchmark
database: multiple LLM agents independently generate SQL, and disagreement among their
executions signals potential annotation errors. {SAR-Agent~\cite{jin2025pervasive} belongs to the same paradigm: it iteratively generates targeted verification queries, executes them on the database, and distills the inspected results into a per-task verdict through a single agent's sequential reasoning.} \textbf{SQL-difference-based methods} such as SQLDriller~\cite{yang2025sqldriller} instead
generate structurally diverse SQL candidates {that represent alternative plausible logical formulations for answering the natural-language question}, synthesize counterexample database instances that
distinguish non-equivalent queries, and prompt an LLM to ``execute'' the natural-language question
on each instance to identify the \ignorespaces {best-aligned} candidate.

\ignorespaces  {The two paradigms rely on different evidence. Execution-based methods use query outcomes on the benchmark database, allowing \system and SAR-Agent to expose incorrect SQL, data inconsistencies, and query ambiguity without assuming that the question or schema is correct. Their signal remains instance-dependent: an erroneous annotation is hidden when it returns the expected result on that database. SQL-difference-based methods instead synthesize counterexamples from structural differences among candidates, exposing latent SQL errors that the current data hides. In exchange, they assume that the question and schema are correct and cover only pure SQL errors.}

\paragraph{Evaluation setup.}
{We compare \system (Frontier and Legacy), SAR-Agent, and SQLDriller on the full \minibench and a targeted \texttt{LIMIT}/\texttt{DISTINCT} error set.} \ignorespaces  \texttt{LIMIT}/\texttt{DISTINCT} errors are the class most favorable to SQL-difference-based methods: a gold SQL that omits \texttt{DISTINCT} or includes a spurious \texttt{LIMIT} may produce correct results on the benchmark database, leaving no execution disagreement for agents to detect, while SQLDriller's synthesized instances are designed to expose exactly these latent structural differences. We identify 71 queries where manual correction contains a \texttt{LIMIT}/\texttt{DISTINCT} change in \minibench. Because all 71 tasks are errors, every system has precision 1.0.

\paragraph{\ignorespaces  {Overall detection performance.}}
Tables~\ref{tab:sqldriller_full} and~\ref{tab:ld_comparison} report the results. {On \minibench, Frontier leads (F1\,=\,0.9194), while Legacy (0.8412) remains comparable to SAR-Agent (0.8389) and exceeds SQLDriller (0.6456). Legacy also operates at substantially lower cost and latency (Table~\ref{tab:cost_latency}).}\ \ignorespaces  {On the \texttt{LIMIT}/\texttt{DISTINCT} set, Frontier detects 65 errors, SAR-Agent 58, Legacy 57, and SQLDriller 41.} SQLDriller's lower recall reflects
inherent pipeline constraints: detection requires the base model to generate at least one
structurally correct alternative among 10 candidates, and the downstream NL-execution step
introduces additional variability from LLM reasoning errors.

\begin{table}[t]
\centering
\caption{Detection performance of \system and existing auditing systems on \minibench.}
\label{tab:sqldriller_full}
\small
\setlength{\tabcolsep}{4pt}
\begin{tabular}{@{}lrrr@{}}
\toprule
\textbf{System} & \textbf{F1} & \textbf{Precision} & \textbf{Recall} \\
\midrule
{\textbf{\system (Frontier)}} & {\textbf{0.9194}} & {\textbf{0.9112}} & {\textbf{0.9277}} \\
\textbf{\system ({Legacy})}  & \textbf{0.8412} & \textbf{0.8218} & \textbf{0.8614} \\
{SAR-Agent (o3)} & {0.8389} & {0.8466} & {0.8313} \\
SQLDriller                & 0.6456 & 0.6800 & 0.6145 \\
\bottomrule
\end{tabular}
\end{table}

\begin{table}[t]
\centering
\caption{Detection performance of \system and existing auditing systems on the \texttt{LIMIT}/\texttt{DISTINCT} error set ($n$\,=\,71).}
\label{tab:ld_comparison}
\small
\setlength{\tabcolsep}{4pt}
\begin{tabular}{@{}lrrrr@{}}
\toprule
\textbf{System} & \textbf{F1} & \textbf{Recall} & \textbf{TP} & \textbf{FN} \\
\midrule
{\textbf{\system (Frontier)}} & {\textbf{0.9559}} & {\textbf{0.9155}} & {\textbf{65}} & {\textbf{6}} \\
\textbf{\system ({Legacy})} & 0.8906 & 0.8028 & 57 & 14 \\
{SAR-Agent (o3)} & {0.8992} & {0.8169} & {58} & {13} \\
SQLDriller & 0.7321 & 0.5775 & 41 & 30 \\
\bottomrule
\end{tabular}
\end{table}

{\paragraph{Per-error-type comparison.}
Table~\ref{tab:error_type_compare} shows that Frontier leads or ties every category except \emph{Data Inconsistency} and has the highest specificity (0.886). Its advantage over SAR-Agent concentrates in \emph{Incorrect SQL} and \emph{Query Ambiguity}, where execution disagreement exposes wrong SQL and divergent interpretations. Legacy matches or beats SAR-Agent in every category but one; SAR-Agent leads only on \emph{Data Inconsistency}, where live-database verification can expose dirty data missed by shared execution results. SQLDriller trails overall.}

\begin{table}[b]
\centering
\caption{{Per-error-type recall on \minibench; the last row is specificity. Frontier and Legacy denote \system configurations; SAR and SQLD denote SAR-Agent and SQLDriller.}}
\label{tab:error_type_compare}
\footnotesize
\setlength{\tabcolsep}{2pt}

\begin{tabular}{@{}lrrrrr@{}}
\toprule
\textbf{Error Type} & \textbf{n} & \textbf{Frontier} & \textbf{Legacy} & \textbf{SAR} & \textbf{SQLD} \\
\midrule
Incorrect SQL & 134 & \textbf{0.95} (127) & 0.88 (118) & 0.85 (114) & 0.63 (84) \\
Query Ambiguity & 14 & \textbf{1.00} (14) & 0.86 (12) & 0.79 (11) & 0.64 (9) \\
Misinterp.\ Domain Sem. & 8 & \textbf{0.75} (6) & \textbf{0.75} (6) & \textbf{0.75} (6) & \textbf{0.75} (6) \\
Data Inconsistency & 7 & 0.57 (4) & 0.57 (4) & \textbf{0.71} (5) & 0.43 (3) \\
Incorrect Evidence & 3 & \textbf{1.00} (3) & \textbf{1.00} (3) & 0.67 (2) & 0.00 (0) \\
\midrule
All error types & 166 & \textbf{0.93} (154) & 0.86 (143) & 0.83 (138) & 0.61 (102) \\
Correct (specificity) & 132 & \textbf{0.89} (117) & 0.77 (101) & 0.81 (107) & 0.64 (84) \\
\bottomrule
\end{tabular}
\end{table}
{\paragraph{Cost and latency.}
Table~\ref{tab:cost_latency} reports cost and latency on 50 random \bird tasks and projects total cost to all 1{,}534 tasks. For SQLDriller's cost profiling, we use GPT-4.1, the default model of its released artifact. SAR-Agent uses its default o3 model. Frontier provides the strongest detection while remaining 1.4$\times$ cheaper and 9.8$\times$ faster than SQLDriller and 2.0$\times$ cheaper and 2.6$\times$ faster than SAR-Agent. Using lower-cost, non-reasoning models, Legacy matches SAR-Agent and outperforms SQLDriller at \$0.0039 and 2.6\,s per task. SQLDriller's 14.6 candidate-generation and NL-execution calls drive its API cost, while counterexample synthesis makes it the slowest system. SAR-Agent's longer iterative reasoning calls produce the highest token usage and cost. Majority vote, agent stacking, and weak labels reuse the same agent calls, while \system's downstream phases add no API calls and only a one-time ${\approx}50$\,s CPU pass.}

\begin{table}[t]
\centering
\caption{{Cost and latency on \bird. Values are per task except total cost, projected to all 1{,}534 tasks from the 50-task sample.}}
\label{tab:cost_latency}
\small
\setlength{\tabcolsep}{2pt}

\begin{tabular}{@{}lrrrrr@{}}
\toprule
\textbf{System} & \textbf{Calls} & \textbf{Total Tokens} & \textbf{Cost (\$)} & \textbf{Latency} & \textbf{Total \$} \\
\midrule
\textbf{\system (Frontier)} & 4.0 & 5{,}499 & 0.0332 & 15.3\,s & 50.89 \\
\textbf{\system (Legacy)} & 4.0 & \textbf{3{,}402} & \textbf{0.0039} & \textbf{2.6\,s} & \textbf{6.04} \\
SQLDriller (GPT-4.1) & 14.6 & 18{,}100 & 0.0449 & 149.7\,s & 68.91 \\
SAR-Agent (o3) & 7.7 & 36{,}966 & 0.0680 & 39.9\,s & 104.38 \\
\bottomrule
\end{tabular}
\end{table}

\finding{2}{\ignorespaces  {\system outperforms SQLDriller and SAR-Agent on both evaluation sets. The Frontier configuration provides the strongest detection performance, while the Legacy configuration preserves comparable performance at substantially lower cost and latency.}}

\subsection{Benchmark-Wide Error Rate Estimation}\label{sec:true_error}

We apply \system under the frontier agent configuration to \ignorespaces  {two versions of the \bird development set~\cite{li2023bird} and the \spider development set~\cite{yu2018spider}} and report both the raw detected error rate and a calibrated estimate of the true error rate.

\paragraph{Calibration method.}

The raw detected rate \ignorespaces {$e_{\mathrm{det}}$ (the fraction of tasks that \system flags on the full development set)} conflates true positives with false positives. \ignorespaces  {For each benchmark version, we use the corresponding manually labeled 200-task random sample to measure \system's true positive rate (TPR) and false positive rate (FPR) and estimate the unknown true benchmark error rate $e_{\mathrm{true}}$.} Under the standard binary-classifier confusion model {(the classical Rogan--Gladen correction~\cite{rogan1978estimating})}, \ignorespaces  {$e_{\mathrm{det}} = \text{TPR} \cdot e_{\mathrm{true}} + \text{FPR} \cdot (1 - e_{\mathrm{true}})$. Solving for $e_{\mathrm{true}}$ yields $\widehat{e}_{\mathrm{true}} = (e_{\mathrm{det}} - \text{FPR}) / (\text{TPR} - \text{FPR})$. The resulting $\widehat{e}_{\mathrm{true}}$ is the calibrated error rate.} \ignorespaces  {We construct its 95\% confidence interval using the delta method.} Table~\ref{tab:error_rate} reports results.

\begin{table}[t]
\centering
\caption{Benchmark-wide error rate estimation. \ignorespaces  {Each row uses TPR and FPR from a manually labeled 200-task sample of the corresponding benchmark version.}}
\label{tab:error_rate}
\small
\setlength{\tabcolsep}{3pt}
\begin{tabular}{@{}lccc@{}}
\toprule
\textbf{Benchmark} & \textbf{Det.\ Rate} & \textbf{Est.\ True Rate} & \textbf{95\% CI} \\
\midrule
BIRD dev (SQLDriller)  & 42.8\% (656/1534) & $\approx$39\% & [33\%, 45\%] \\
BIRD dev (Nov.\ 2025)  & 37.2\% (571/1534) &

  {$\approx$37\%}

&

  {[30\%, 43\%]}

\\
Spider dev             & 31.4\% (325/1034) &

  {$\approx$27\%}

&

  {[21\%, 33\%]}

\\
\bottomrule
\end{tabular}
\end{table}

\paragraph{BIRD development set.}
{On the manually labeled 200-task sample from the SQLDriller-version BIRD development set, SALUS achieves F1 = 0.8552, TPR = 0.91, and FPR = 0.11. Calibrating the full-set detected rate with these sample-specific rates estimates an annotation error rate of approximately 39\% (95\% CI: [33\%, 45\%]). For the November 2025 release, we re-evaluate the same 200 tasks under the same labeling protocol described in Section~\ref{sec:labeling}; SALUS achieves F1 = 0.8429, TPR = 0.86, and FPR = 0.09. The corresponding calibrated error-rate estimate is approximately 37\% (95\% CI: [30\%, 43\%]), showing that errors remain pervasive after the update.}

\paragraph{Spider development set.}
\ignorespaces  {On the manually labeled 200-task \spider sample, \system achieves F1\,=\,0.8750, with TPR\,=\,0.93 and FPR\,=\,0.09. The corresponding calibrated error-rate estimate is approximately 27\% (95\% CI: [21\%, 33\%]).}

\finding{3}{Annotation errors remain pervasive: approximately \ignorespaces {37\%} in the November 2025 \bird update and \ignorespaces {27\% in \spider after benchmark-specific calibration}. Even after the \bird team's review, a substantial number of errors persist.}

\subsection{Leaderboard Trustworthiness}
\label{sec:leaderboard}

NL-to-SQL benchmarks usually maintain a public leaderboard \ignorespaces {that ranks systems by execution accuracy on both the development and hidden test sets.} For the BIRD development set, we have an estimated error rate of \ignorespaces {37\%}; the test set labels are hidden and their quality is unknown.
\ignorespaces {Using the BIRD benchmark as an example,} we first analyze how such development set error rate affects the trustworthiness of the
development ranking, then infer implications for the test ranking.

For a pair of systems with reported development accuracies $a_v$ and $a_k$ on $N = 1{,}534$
queries, correcting a single flawed query changes each system's accuracy by at most $\pm 1/N$,
producing a maximum pairwise gap swing of $\pm 2/N$. Under the worst-case assumption that all
\ignorespaces {$e = 564$ estimated flawed queries (37\%)} produce such maximal swings independently, the standard
deviation of the total gap perturbation is $\sigma_{\mathrm{gap}} \le 2\sqrt{e}/N \approx \ignorespaces {3.10\%}$. A one-sided 95\% confidence threshold then requires an observed gap of at least $g^* = 1.645\,\sigma_{\mathrm{gap}} \approx \ignorespaces {5.09\%}$ (derivation in Appendix~\ref{app:variance} and~\ref{app:gap_prob}).

We applied this analysis to the top 30 systems ranked by development set accuracy, \ignorespaces as shown in Figure~\ref{fig:bird_leaderboard_trust}. No pairwise comparison between the rank-1 system and any of the \ignorespaces {next 20} systems exceeds the 95\% threshold. In other words, given the estimated \ignorespaces {37\%} error rate, the
development set ranking cannot reliably distinguish the top-ranked system from the \ignorespaces {next 20}. This analysis assumes that corrections
to flawed queries affect each system independently. In practice, systems with shared architectures are likely to respond similarly to the same corrections. This positive correlation reduces the effective variance, making $\sigma_{\mathrm{gap}} \le 2\sqrt{e}/N$ an upper bound. Therefore, we are making a conservative analysis: the actual indistinguishability region may be narrower than reported.

The development ranking is only a proxy; the official BIRD leaderboard is determined by the hidden test set. We therefore ask whether the test ranking behaves like an independent and substantially cleaner reference, or whether it may reflect similar benchmark-specific artifacts.

We examine this question through rank correlation. The standard Kendall's $\tau$ between the observed development ranking and the hidden test ranking for the top 30 systems by hidden-test accuracy is $\tau = 0.664$. Because Kendall's $\tau$ is a U-statistic, linearity of expectation under our pairwise noise model allows us to compute an expected Kendall's $\tau$ ($E[\tau]$). This expectation mechanically down-weights pairs whose score gaps fall within the noise margin, as their true ordering is highly uncertain. Under our estimated \ignorespaces {37\%} error rate, we obtain $E[\tau] = \ignorespaces {0.41}$ (derivation in Appendix~\ref{app:ktau}).

If the hidden test set were a pristine oracle of true NL-to-SQL capability, pairs that are statistically indistinguishable on the development set should have a near-random chance of maintaining their relative order on the test set. Consequently, the observed $\tau$ should closely track the down-weighted $E[\tau]$. Instead, the observed $\tau$ ($0.664$) is substantially higher than $E[\tau]$ ($\ignorespaces {0.41}$). This discrepancy indicates that small-gap pairs are concordant between the development and test sets far more often than chance would predict.

\begin{table}[b]
\centering
\caption{Agent--gold and inter-agent disagreement rates on the BIRD development set. Only the highest and lowest rates are shown for each category;}
\label{tab:agent_disagreement}
\small
\begin{tabular}{lc}
\toprule
\textbf{Statistic} & \textbf{Rate} \\
\midrule
\multicolumn{2}{l}{\textit{Agent vs.\ gold disagreement}} \\
GPT-5 vs.\ Gold (highest)          & 45.4\% \\
Opus~4.6 vs.\ Gold (lowest)        & 36.7\% \\
\midrule
\multicolumn{2}{l}{\textit{Pairwise disagreement between agents}} \\
GPT-5 vs.\ GPT-4o (highest)        & 37.1\% \\
Opus~4.6 vs.\ Gemini~2.5~Pro (lowest) & 30.4\% \\
\midrule
\multicolumn{2}{l}{\textit{Collective statistics}} \\
All 4 agents mutually agree         & 50.3\% \\
Majority of valid agents disagrees with gold & 46.9\% \\
\bottomrule
\end{tabular}
\end{table}

This excess concordance on statistically noisy pairs is consistent with the hypothesis that development and test sets share similar
annotation artifacts, so that systems rewarded by one split tend to be rewarded by the other. Alternative explanations include our noise
model overestimating development-side uncertainty, split-specific variance, or distribution shift. Even under these more conservative interpretations, however, we cannot claim that the hidden test leaderboard provides a statistically reliable ordering of state-of-the-art systems. Until benchmark label quality is better established across all splits, leaderboard differences at the top should be interpreted with caution.

\finding{4}{Under the estimated \ignorespaces {37\%} error rate, pairwise accuracy gaps below \ignorespaces {5.09\%} are statistically indistinguishable at the 95\% confidence level, and the top-ranked system on the \bird development set cannot be reliably separated from the \ignorespaces {next 20}. If the hidden test set were a clean independent reference, the observed rank correlation should approximate $E[\tau] = \ignorespaces {0.41}$; instead, the observed $\tau = 0.664$ indicates that small-gap pairs are concordant far more often than chance predicts, consistent with shared annotation artifacts across both splits.}

\subsection{System Analysis}\label{sec:system_eval}

\system achieves strong detection performance, and its predictions reveal substantial benchmark noise. We now analyze the internal behavior of the pipeline to characterize the properties that drive this performance. These statistics are computed over SQLDriller \bird development set ($n=1{,}534$)~\cite{yang2025sqldriller}, using unlabeled data without dependence on ground-truth annotations.

\subsubsection{Multi-Agent Agreement}\label{sec:agent_agreement}

\system collects candidate SQL from four frontier agents and executes them against each database.
Table~\ref{tab:agent_disagreement} summarizes the resulting disagreement patterns.

Agent disagreement with gold annotations is frequent (36.7\% to 45.4\%) and potentially informative as a detection signal. Pairwise disagreement between agents ranges from 30.4\% to 37.1\%, confirming that the agents are not strongly correlated. Their outputs are also stable across reruns. At the collective level, all four agents mutually agree on only 50.3\% of tasks, meaning that on nearly half of all tasks the agents diverge from one another and \system is designed to determine which agent actually returns the correct verdict. Furthermore, a majority of valid agents disagree with gold in 46.9\% of tasks. Given that these are four of the most capable frontier language models available, such widespread collective disagreement with the benchmark annotations suggests that a substantial fraction of gold SQL queries are likely erroneous. We also reran each agent eight times on 15 randomly sampled queries (120 verdicts per agent); disagreement with the original verdict was low: 3.33\% for Claude, 4.17\% for GPT-4o, 7.50\% for GPT-5, and 8.33\% for Gemini, showing that the agents' verdicts are stable across repeated runs.

\subsubsection{{LF Correlations}}\label{sec:lf_independence}
{Effective aggregation benefits when the LFs provide complementary evidence across signal sources. {}We compute the mean pairwise Pearson correlation of non-abstaining votes among LF pairs. Within-category pairs exhibit a mean correlation of 0.567, while cross-category pairs have a mean correlation of 0.081, a sevenfold reduction confirming that the three signal sources capture different facets of annotation quality. This low cross-source dependence is the condition under which the generative label model can effectively denoise conflicting votes.}

{The correlation analysis concerns the three task-level LF families; the meta LF instead summarizes their non-abstaining votes, densifying rows because 30 of the 44 LFs are unipolar and frequently abstain. Refitting without it changes predicted error probabilities by 0.007 on average and 4 of 1,534 weak labels, showing that this aggregation does not materially inflate the label-model output.}

\subsubsection{{LF Signal-source Ablation}}\label{sec:lf_ablation}

\begin{table}[t]
\centering
\caption{Ablation of LF \ignorespaces groups on \minibench{, including every valid multi-group removal}.}
\label{tab:lf_ablation}
\small
\setlength{\tabcolsep}{4pt}
\begin{tabular}{@{}lcc@{}}
\toprule
\textbf{
  {Ablation Configuration}
} & \textbf{F1} & \textbf{$\Delta$F1} \\
\midrule
\multicolumn{3}{@{}l}{{\textit{Single-group removals}}} \\
None (full system)       & 0.9194 & --- \\
$-$ Structural signals   & 0.8758 & $-$0.0436 \\
$-$ Execution signals    & 0.8829 & $-$0.0365 \\
$-$ Meta LF              & 0.8976 & $-$0.0218 \\
$-$ Intent signals       & 0.9024 & $-$0.0170 \\
\midrule
\multicolumn{3}{@{}l}{{\textit{Two-group removals}}} \\
{$-$ Execution $-$ Structural} & {0.6467} & {$-$0.2727} \\
{$-$ Execution $-$ Intent} & {0.8715} & {$-$0.0479} \\
{$-$ Execution $-$ Meta} & {0.8896} & {$-$0.0298} \\
{$-$ Structural $-$ Intent} & {0.8909} & {$-$0.0285} \\
{$-$ Structural $-$ Meta} & {0.8758} & {$-$0.0436} \\
{$-$ Intent $-$ Meta} & {0.8836} & {$-$0.0358} \\
\midrule
\multicolumn{3}{@{}l}{{\textit{Three-group removals (remaining family shown)}}} \\
{Intent signals only} & {0.6052} & {$-$0.3142} \\
{Structural signals only} & {0.8882} & {$-$0.0312} \\
{Execution signals only} & {0.8709} & {$-$0.0485} \\
\bottomrule
\end{tabular}
\end{table}

Table~\ref{tab:lf_ablation} reports detection F1 after removing each \ignorespaces {LF} group and retraining the pipeline. {Every single-group removal lowers F1: structural by 0.0436, execution by 0.0365, and intent by 0.0170.} \ignorespaces  {Removing the meta LF lowers F1 by 0.0218, confirming that its aggregate consensus signal contributes to final detection performance.} \ignorespaces  {Together with the low cross-source correlations in Section~\ref{sec:lf_independence}, these losses show that the groups provide complementary evidence. Structural signals catch missing joins or aggregation mismatches; execution signals compare results; intent signals check question--SQL alignment, while the meta LF supplies a consensus signal when few LFs vote.}

{The multi-group removals reveal compensatory coverage. Using only structural or execution signals retains F1 of 0.8882 and 0.8709, respectively, whereas removing both sources reduces F1 from 0.9194 to 0.6467, far beyond their summed single-group loss. Thus, either primary source provides fallback evidence when LFs from the other source abstain or miss an error, while intent signals complement these paths by targeting rarer question--SQL alignment errors.}

\subsubsection{Decision Plane and Feature-Conditioned Routing}\label{sec:feature_analysis}

\begin{table}[t]
\centering
\caption{Top features in the depth-2 CART reliability model for each frontier agent, ranked by Gini importance.}
\label{tab:feature_importance}
\small
\begin{tabular}{ll}
\toprule
\textbf{Agent} & \textbf{Top Features} \\
\midrule
GPT-5 &
\texttt{has\_subquery}, \texttt{agg\_count} \\
GPT-4o &
\texttt{join\_count}, \texttt{table\_count} \\
Opus~4.6 &
\texttt{join\_count}, \texttt{agg\_count}, \texttt{has\_distinct} \\
Gemini~2.5~Pro &
\texttt{table\_count}, \texttt{cardinality} \\
\bottomrule
\end{tabular}
\end{table}

The selected configuration uses depth-2 CART paired with score fusion. {On \minibench, it achieves F1\,=\,0.9194 versus 0.9042 for depth-4 CART and 0.9059 for XGBoost, indicating that additional capacity does not improve generalization from the high-confidence weak labels.} Because the trees are shallow, Gini importance directly reflects each feature's contribution to separating regions of high vs.\ low agent reliability. Table~\ref{tab:feature_importance} shows that each agent relies on a distinct set of top features: GPT-5 splits on subquery presence and aggregation count; GPT-4o on join count and table count; Opus~4.6 on join count, aggregation count, and distinct usage; and Gemini~2.5~Pro on table count and result cardinality. These structurally distinct reliable regions confirm that feature-conditioned routing extracts complementary agent strengths that uniform voting cannot, directly addressing the challenge of agent fallibility. The score fusion strategy further incorporates the fallback task-level probability as a continuous feature on every query, providing a learned safety signal even when agents produce valid but unreliable verdicts.

{Agent stacking replaces the SQL-feature-conditioned decision plane with a single global mapping over agent verdicts (Section~\ref{sec:baselines}). We therefore add a study to demonstrate the benefit of the decision plane over this alternative by testing whether its global mapping can represent the decision plane's task-specific routing. All 16 four-agent verdict patterns occur in \minibench, and 11 contain both correct and erroneous annotations, accounting for 275 of the 298 tasks. A global mapping must assign the same label to every task with the same verdict pattern. For this diagnostic, we exhaustively evaluate all $65{,}536$ mappings using the 298 manual correctness labels, giving agent stacking an optimistic evaluation-only ceiling. Its maximum F1 is 0.9112, below \system's 0.9194. BIRD development-set tasks with IDs 847 and 18 illustrate this mapping rule's limitation: GPT-5 and Claude Opus 4.6 vote correct for both reference annotations, while GPT-4o and Gemini 2.5 Pro vote erroneous. However, task 847's reference SQL has a higher table count than task 18's (3 versus 2). Consistent with Table~\ref{tab:feature_importance}, where table count is a top feature for GPT-4o, this difference leads \system to trust GPT-4o's disagreement on erroneous task 847 but downweight the same verdict on correct task 18, correctly judging both tasks. A global mapping rule cannot make this distinction.}

Together, these analyses show how \system converts noisy multi-agent output into reliable benchmark audits.

\finding{5}{Our system-wise experiments show that each LF signal source captures a distinct facet of annotation quality, that removing any source degrades detection, and that per-agent reliability models learn to trust different agents in different regions of the SQL feature space. \ignorespaces }

\section{Discussion}
\subsection{Beyond Detection to Benchmark Correction}\label{sec:discussion_correction}

Detection identifies which benchmark annotations are likely wrong; the natural next step is to
\ignorespaces  {make} corrections. \ignorespaces 
{For example,} SQLDriller~\cite{yang2025sqldriller} {automatically} selects the SQL candidate with the highest NL-execution
consistency score from 10 model-generated alternatives, provided that it scores strictly higher than the reference SQL. \ignorespaces  \system includes a lightweight {automatic} correction mechanism: when a detected error has a majority-agreement cluster of at least three valid agents, the system proposes the SQL from
the highest-reliability agent in that cluster. \ignorespaces  {Table~\ref{tab:correction_compare} compares both mechanisms on the 134 \textit{Incorrect SQL} errors. \system produces more correct fixes than SQLDriller (59 vs.\ 35), with a higher end-to-end fix rate (44.0\% vs.\ 26.1\%). Reliable detection is necessary but not sufficient for correction: our mechanism fixes only 44.0\% of the \textit{Incorrect SQL} errors. Beyond this category, 19.3\% of \minibench errors involve question ambiguity, evidence, domain semantics, or data, for which SQL substitution alone may be insufficient (Section~\ref{sec:error_stats}). We therefore leave robust correction to future work.}

\begin{table}[t]
\centering
\caption{{Automatic correction of the 134 \textit{Incorrect SQL} errors; correctness requires execution match with the manual correction.}}
\label{tab:correction_compare}
\small

\begin{tabular}{@{}lrr@{}}
\toprule
\textbf{Quantity} & \textbf{\system} & \textbf{SQLDriller} \\
\midrule
Errors detected (of 134) & \textbf{127 (94.8\%)} & 84 (62.7\%) \\
Usable correction proposals & 75 & 69 \\
Execution-match with manual fix & \textbf{59} & 35 \\
Conditional accuracy & \textbf{78.7\%} & 50.7\% \\
End-to-end correct fix rate & \textbf{44.0\%} & 26.1\% \\
\bottomrule
\end{tabular}
\end{table}

Several directions could improve \system's correction quality: learning supervised agent selection from a ground-truth correction set rather than agreement-cluster heuristics; using stronger NL-to-SQL systems, such as multi-step agentic pipelines, to increase the chance that candidates contain a correct alternative; and localizing error types, such as wrong joins or missing aggregations, as structured feedback to annotators. In all cases, correction quality is bounded by detection accuracy: a system cannot fix what it fails to identify. Robust detection is therefore essential for automated correction.

\subsection{Broader Applications}\label{sec:discussion_applications}
\system can extend beyond BIRD and Spider to broader auditing needs.

\paragraph{Real-time auditing.}
A configured \system instance can serve as an online quality gate for incoming annotations, generating agent answers, extracting features, and returning an immediate detection result. As tasks accumulate, the system can periodically retrain its decision plane to improve accuracy.

\paragraph{More complex benchmark settings.}
{\system is source agnostic: it consumes candidate SQL and execution verdicts rather than depending on a particular model family or provider, so an open-weight model enters through the same interface.} Therefore, it can in principle be applied to more challenging benchmarks where single-prompt LLM agents are insufficient. Spider~2.0~\cite{spider2}, for example, features enterprise-scale schemas with over 3,000 columns and multi-step workflows where even frontier LLMs achieve success rates below 30\%~\cite{spider2}. At this difficulty level, more advanced multi-agent pipelines would be required as source agents. \ignorespaces  {We demonstrate this property by adding SQLDriller as a fifth signal source without changing the architecture. This configuration obtains F1\,=\,0.9102 on \minibench, lower than the four-source result of 0.9194 because a compatible additional source does not necessarily improve accuracy.} However, on more challenging benchmarks, most top-performing systems are proprietary, limiting access to diverse agent signals and therefore requiring broader collaboration between benchmark maintainers and system developers.

\paragraph{Transfer to other benchmark auditing tasks.}
The multi-agent weak labeling architecture is not specific to NL-to-SQL and can transfer to any domain where benchmark annotations may contain errors. For software engineering benchmarks such as LiveCodeBench~\cite{jain2024livecodebenchholisticcontaminationfree} and SWE-bench~\cite{jimenez2024swebenchlanguagemodelsresolve}, where weak test suites accept incorrect patches and inflate resolution rates, labeling functions can check whether multiple agents converge on the same fix location and whether their patches produce identical behavior under differential testing, with reliability estimated from patch complexity features such as the number of files and functions modified. For mathematical reasoning benchmarks such as RealMath~\cite{zhang2025realmathcontinuousbenchmarkevaluating}, where the automated theorem-to-QA extraction pipeline introduces annotation noise, labeling functions can check whether agents converge on an alternative numerical answer that disagrees with the reference, with reliability estimated from problem features such as mathematical domain and the number of reasoning steps. In both cases, the weak supervision framework and decision plane ensemble remain unchanged; only the domain-specific labeling functions and reliability criteria require adaptation.

\section{Conclusion}
In this paper, we presented \system, an automated and benchmark-agnostic framework for detecting annotation errors in NL-to-SQL benchmarks through weak supervision over multi-agent SQL outputs. By combining multiple labeling functions, a generative label model, and a feature-conditioned decision plane, \system achieves strong end-to-end detection performance on our manually verified benchmark \minibench, reaching an F1 of 0.9194 and outperforming all baselines. When applied at a benchmark scale, \system estimates that annotation errors remain substantial even after recent cleanup efforts: approximately \ignorespaces {37\%} in the November 2025 BIRD update and \ignorespaces {27\%} in Spider. These errors directly affect evaluation reliability: on BIRD, pairwise rankings with development accuracy gaps within \ignorespaces {5.09\%} cannot be distinguished, making much of the current leaderboard statistically unreliable. {Therefore, for benchmark maintainers, we suggest four concrete steps: (1) run automated auditing such as \system as a release gate, which at \$0.0332 per task is repeatable on every update; (2) prioritize flagged tasks for multi-annotator adjudication rather than re-reviewing entire benchmarks; (3) maintain a verified-clean subset for leaderboard comparison; and (4) report ranking uncertainty alongside scores.} Taken together, our results establish benchmark auditing as a first-class problem in NL-to-SQL evaluation and show that scalable systems such as \system can support cleaner benchmarks.

\bibliographystyle{ACM-Reference-Format}
\bibliography{references}

\clearpage
\appendix
\begingroup
\providecommand{\dep}[1]{\ignorespaces}
\providecommand{\revisionA}[2]{#2}
\section{Labeling Function Configurations}
\label{app:lf}

This appendix documents the labeling functions used in the final paper pipeline.
The final configuration contains 43
task-level labeling functions and one meta labeling function.

\subsection{Execution signals.}
\paragraph{Result-set signals.}
These signals compare the result set $R(s_i, D_i)$ produced by the benchmark
SQL $s_i$ on database $D_i$ against the execution results produced by the agent
queries. These functions normalize away benign formatting differences such as
whitespace, case, NULL representation, column order, and small floating-point
variation. They capture direct result-set agreement, runtime failures, cardinality
mismatches, subset/superset relations, and numeric-equivalence edge cases such as
percentage-vs-fraction scaling.

\paragraph{Agent Consensus signals.}
Consensus signals do not inspect the natural-language question directly. Instead,
they summarize agreement patterns across the valid agent executions themselves.
In the appendix we name these rules \emph{Agent Majority Agreement} and
\emph{Agent Unanimity Agreement}. These are weak-supervision heuristics used in
the labeling-function layer; they should not be confused with the later Decision
Plane, which learns per-agent reliability scores from SQL features.


\subsection{Structural signals.}
Structural signals compare the SQL structure of $s_i$ against the agent-consensus
SQL using AST-level analysis. Let $\mathit{Tables}(s)$, $\mathit{Aggs}(s)$, and
$\mathit{GroupBy}(s)$ denote the referenced tables, aggregation operators, and
\texttt{GROUP BY} expressions extracted from the AST of query $s$. These rules
detect missing or extra joins, aggregation mismatches, grouping mismatches,
join-type discrepancies, suspicious hard-coded literals, and semantically
equivalent rewrites. We also include one schema-consistency rule that rewards
cases where the benchmark SQL and the consensus SQL reference the same table set,
as well as a family of known-error-pattern rules targeting recurring annotation
mistakes such as aggregation without grouping, date-boundary off-by-one errors,
unguarded division, implicit type coercion, and invalid \texttt{DISTINCT} /
\texttt{ORDER BY} combinations.

Many negative structural rules use an \emph{execution guard}: they vote
\textsc{Incorrect} only when a structural discrepancy is accompanied by an
execution-level mismatch. This reduces noise from harmless syntactic variation.
For example, if the benchmark query and the consensus query differ in join type
but still produce the same result on the current database, the guarded rule
abstains rather than casting an \textsc{Incorrect} vote.

\subsection{Intent signals.}
Intent signals capture failures that may not be visible from execution or SQL
structure alone. These rules check whether the SQL reflects salient cues in the
question, such as whether ``how many'' maps to \texttt{COUNT}, ``average'' maps
to \texttt{AVG}, and ``top $N$'' maps to \texttt{LIMIT}, while still allowing
alternative SQL formulations that are semantically equivalent. Additional rules
check for missing filter literals, omitted \texttt{DISTINCT}, and yes/no
questions that incorrectly return a full table rather than a boolean-style
answer.

\subsection{Meta signal.}
A single meta labeling function computes a majority vote over the non-abstaining
task-level labeling-function outputs. This rule provides an aggregate summary of
the direction of evidence. \dep{Because it is deterministically derived from the
task-level votes, it introduces dependence into the label matrix; we retain it
because label models are typically robust to moderate violations of conditional
independence in practice~\cite{ratner2017snorkel}.}
\revisionA{R1.O5}{Although the meta LF is deterministically derived from the task-level votes,
conditional independence is not required by Snorkel: its generative model can capture pairwise LF
dependencies and avoid double-counting correlated votes~\cite{ratner2017snorkel}.}


\section{Pairwise Gap Perturbation and Expected Rank Correlation}
\label{app:gap_variance_and_tau}

This appendix derives the probabilistic framework used in Section~\ref{sec:leaderboard} to
quantify uncertainty in development set accuracy gaps and its propagation into rank correlation
with the hidden test set.

\subsection{Pairwise Gap Perturbation Variance}
\label{app:variance}

Let $N$ be the development set size and $e$ the estimated number of flawed queries. For two
systems $v$ and $k$ with reported accuracies $a_v$ and $a_k$, define the observed gap
$g_{v,k} = a_v - a_k$. After correcting all flawed queries, the true gap is
\[
g_{v,k}^\star = g_{v,k} + \sum_{i=1}^{e} \delta_i^{(v,k)},
\]
where $\delta_i^{(v,k)}$ is the change in the gap from correcting the $i$-th flawed query.

\paragraph{Per-query bound.}
Correcting a single query changes each system's accuracy by at most $\pm 1/N$, so the pairwise
gap change satisfies $\delta_i^{(v,k)} \in \{-2/N,\, -1/N,\, 0,\, +1/N,\, +2/N\}$ and
$|\delta_i^{(v,k)}| \le 2/N$.

\paragraph{Variance bound.}
Adopting a symmetry assumption $\mathbb{E}[\delta_i^{(v,k)}] = 0$, the per-query variance is
bounded by
\[
\operatorname{Var}(\delta_i^{(v,k)}) = \mathbb{E}\!\left[(\delta_i^{(v,k)})^2\right]
\le \left(\frac{2}{N}\right)^2.
\]
Assuming independence across the $e$ flawed queries, the total perturbation
$\Delta_{v,k} = \sum_{i=1}^e \delta_i^{(v,k)}$ has variance
\[
\operatorname{Var}(\Delta_{v,k}) = \sum_{i=1}^{e} \operatorname{Var}(\delta_i^{(v,k)})
\le e \cdot \frac{4}{N^2},
\]
yielding the standard deviation bound
\[
\sigma_{\mathrm{gap}} \le \frac{2\sqrt{e}}{N}.
\]
This bound depends only on $e$ and $N$ and is therefore identical for all system pairs.

\subsection{Outperformance Probability}
\label{app:gap_prob}

Since $\Delta_{v,k}$ is a sum of $e$ independent, bounded, zero-mean random variables, the
Central Limit Theorem gives $\Delta_{v,k} \approx \mathcal{N}(0, \sigma_{\mathrm{gap}}^2)$.
The corrected gap is then distributed as
\[
g_{v,k}^\star \approx \mathcal{N}(g_{v,k},\; \sigma_{\mathrm{gap}}^2).
\]
The probability that system $v$ truly outperforms system $k$ on a corrected development set is
\[
P(v >_{\mathrm{dev}} k) = P(g_{v,k}^\star > 0)
= P\!\left(\frac{\Delta_{v,k}}{\sigma_{\mathrm{gap}}} > \frac{-g_{v,k}}{\sigma_{\mathrm{gap}}}\right)
= \Phi\!\left(\frac{g_{v,k}}{\sigma_{\mathrm{gap}}}\right),
\]
where $\Phi(\cdot)$ is the standard normal CDF. A one-sided 95\% confidence threshold requires
$g_{v,k} \ge 1.645\,\sigma_{\mathrm{gap}}$.

\subsection{Expected Kendall's $\tau$}
\label{app:ktau}

Standard Kendall's $\tau$ treats reported development scores as deterministic. To incorporate
annotation uncertainty, we derive an Expected Kendall's $\tau$ using the fact that $\tau$ is a
U-statistic, which permits application of linearity of expectation
over independent pairwise comparisons.

Let $M$ systems be ranked by the hidden test set. For each pair $(i, j)$ with $i$ ranked higher
on the test set, define the sign variable
\[
s_{i,j} =
\begin{cases}
+1 & \text{concordant}, \\
-1 & \text{discordant}.
\end{cases}
\]
where concordance means the development ranking agrees with the test ranking for the pair,
and discordance means it disagrees. Standard Kendall's $\tau$ is
\[
\tau = \frac{1}{\binom{M}{2}} \sum_{i < j} s_{i,j}.
\]
Under our noise model, each deterministic $s_{i,j}$ is replaced by its expectation. Since $i$ is
ranked higher on the test set, concordance requires $i >_{\mathrm{dev}} j$, giving
\[
\mathbb{E}[s_{i,j}] = (+1) \cdot P(i >_{\mathrm{dev}} j) + (-1) \cdot (1 - P(i >_{\mathrm{dev}} j))
= 2\,P(i >_{\mathrm{dev}} j) - 1.
\]
The Expected Kendall's $\tau$ is then
\[
E[\tau] = \frac{1}{\binom{M}{2}} \sum_{i < j} \bigl(2\,P(i >_{\mathrm{dev}} j) - 1\bigr).
\]
When a pair's gap falls within the noise margin ($P \approx 0.5$), its contribution approaches
zero, effectively neutralizing statistically insignificant comparisons.

\endgroup

\end{document}